\documentclass[aps,twocolumn,showpacs,superscriptaddress,pra,10pt]{revtex4-1}
\usepackage{graphicx}
\usepackage{bm}
\usepackage{amsmath}
\usepackage{amsfonts}

\usepackage{tikz}
\usetikzlibrary{intersections,shapes.arrows}
\usetikzlibrary{shapes.geometric,positioning}
\usetikzlibrary{shapes}
\usetikzlibrary{decorations}
\usetikzlibrary{patterns,arrows,decorations.pathreplacing}
\usepackage{hyperref} 

\newcommand{\energycommand}{E}
\begin{document}

\author{M. Motta}
\affiliation{Department of Physics, College of William and Mary, Williamsburg, Virginia 23187-8795, USA}

\author{E. Vitali}
\affiliation{Department of Physics, College of William and Mary, Williamsburg, Virginia 23187-8795, USA}

\author{M. Rossi}
\affiliation{Scuola Normale Superiore, Piazza dei Cavalieri 7, I-56126 Pisa, Italy}
\affiliation{International Center for Theoretical Physics (ICTP), Strada Costiera 11, I-34154 Trieste, Italy}

\author{D. E. Galli}
\affiliation{Dipartimento di Fisica, Universit\`a degli Studi di Milano, via Celoria 16, I-20133 Milano, Italy}

\author{G. Bertaina}
\affiliation{Dipartimento di Fisica, Universit\`a degli Studi di Milano, via Celoria 16, I-20133 Milano, Italy}

\title{Dynamical structure factor of one-dimensional hard rods}

\begin{abstract}
The zero-temperature dynamical structure factor $S(q,\omega)$ of one-dimensional hard rods is computed
using state-of-the-art quantum Monte Carlo and analytic continuation techniques, complemented by a Bethe Ansatz analysis.
As the density increases, $S(q,\omega)$ reveals a crossover from the Tonks-Girardeau gas to a 
quasi-solid regime, along which the low-energy properties are found
in agreement with the nonlinear Luttinger liquid theory. Our quantitative estimate of $S(q,\omega)$ extends beyond the low-energy limit and confirms a theoretical prediction regarding the behavior 
of $S(q,\omega)$ at specific wavevectors $\mathcal{Q}_n=n 2 \pi/a$, where $a$ is the core radius,
resulting from the interplay of the particle-hole boundaries of 
suitably rescaled ideal Fermi gases. We observe significant 
similarities between hard rods and one-dimensional $^4$He at high density,
suggesting that the hard-rods model may provide an accurate description of 
dense one-dimensional liquids of quantum particles interacting through a strongly repulsive, finite-range potential.
\end{abstract}

\maketitle
\section{Introduction}

One-dimensional (1D) quantum systems are subject to intense research, due to their theoretical and experimental peculiarities 
\cite{Giamarchi2004,Cazalilla2011,Imambekov2012}.
On the theoretical side, the reduced dimensionality enhances quantum fluctuations and interaction, giving rise 
to unique phenomena like the nonexistence of Bose-Einstein condensation \cite{HMW,Mermin1966,Coleman1973,Girardeau1960,Pitaevskii1993} 
and the breakdown of Fermi-liquid behavior \cite{Vignale2005}.
On the experimental side, a 1D system is realized when a 3D system is loaded into an elongated optical trap
or confined into a narrow channel, and the transverse motion is frozen to zero-point fluctuations.

Remarkably, several 1D many-body models of considerable conceptual and experimental relevance are exactly solvable
 \cite{Nagamiya1940,LiebLiniger1963,Lieb1963,Sutherland1971,Sutherland1971b,Sutherland1988}, and provide a precious 
support for understanding static and dynamic properties of interacting 1D systems in suitable regimes 
\cite{expgen1,expgen2,expgen3,expdyn1,expdyn2,expdyn3,expdyn4,expsol1,expsol2,expfm1,expfm2,Fabbri2015,Bertaina2016}.

In particular, the behavior of 1D systems with a hard-core repulsive interaction, 
like ${}^4 \mbox{He}$ or other gases adsorbed in carbon nanotubes \cite{Pearce2005,Mercedes2001,Mazzanti2008b,Mazzanti2008a,Bertaina2016}, can 
be understood making the assumption that particles behave like a gas of impenetrable segments
or hard rods (HRs).
Indeed, at high density, the principal effect of a short-range hard-core repulsive interaction is 
volume exclusion. Therefore, a reasonable approximation of the actual microscale behavior of 
the system can be obtained by taking into account the volume exclusion phenomenon only, 
neglecting all other details of the interaction: within this approach, the system is described as an assembly of HRs of a suitable length $a$.

The recognition that volume exclusion is the most important factor in analyzing 
short-range hard-core repulsive interactions in high-density classical systems dates back to the seminal work by van der Waals \cite{Waals1910} 
and Jeans \cite{Jeans1916}. 
It was later recognized \cite{Tonks1936,Hove1952} that the statistical mechanics of a system of classical HRs is exactly solvable.
In 1940 Nagamiya proved \cite{Nagamiya1940} that also a system of quantum HRs is exactly solvable using the Bethe Ansatz technique, and 
imposing a special system of boundary conditions. 
Nagamiya's treatment was later adapted by Sutherland \cite{Sutherland1971b} to the more familiar periodic boundary conditions. 

It is remarkable that local properties of the HR model are independent of the particles being bosons or fermions \cite{Girardeau1960}, since in 1D the hard core interaction creates nodes in bosonic wavefunctions which can be completely mapped to the nodes of fermionic wavefunctions. Only non-local properties differ, such as the momentum distribution \cite{Mazzanti2008b}. 

Even if the eigenfunctions and eigenvalues of the HR model can be determined exactly, so far 
the only systematic way to obtain a complete description of the ground-state correlation 
functions of the model has been the Variational Monte Carlo (VMC) method \cite{Mazzanti2008b,Mazzanti2008a}. Dynamical properties have also been addressed by using the variational Jastrow-Feenberg theory in \cite{Krotscheck1999}.

In the present work, we resort to state-of-the-art projective quantum Monte Carlo (QMC) \cite{Sarsa2000,Galli2003,Patate} 
and analytic continuation \cite{Vitali2010} techniques to compute the dynamical structure factor, 
$S(q,\omega)$, of a single-component system of HRs. This analysis is supported by the Bethe Ansatz solution of the elementary excitations of the model, following \cite{Lieb1963}. 

The dynamical structure factor characterizes the linear response of the system to an external
field which weakly couples to the density. In the context of quantum liquids, it can be probed via inelastic neutron scattering \cite{Cowley1971,Beauvois2016}, while in the ultracold gases field it can be probed with Bragg scattering \cite{Ozeri2005,Fabbri2015}, also implemented via digital micromirror devices \cite{Ha2015}, or cavity-enhanced spontaneous emission \cite{Landig2015}.

While the low-energy properties of $S(q,\omega)$ are universal and can be described by the 
Tomonaga-Luttinger liquid (TLL) theory \cite{Tomonaga1950,Luttinger1963,Mattis1965,Haldane1983,Haldane1983b}
and its recent and remarkable generalization, called the nonlinear TLL theory\cite{Imambekov2009,Imambekov2012},
high-energy properties depend explicitly on the shape of the interaction potential, and lie in a regime 
beyond the reach of those theoretical approaches. Due to such limitation, we rely on QMC to estimate $S(q,\omega)$ for all momenta and energies.

The HR model, and its solution by Bethe Ansatz, is described in Section \ref{sec:model}. 
The methods used to compute the dynamical structure factor are reviewed in Section 
\ref{sec:method}. Results are presented and discussed in Section \ref{sec:results}, and 
conclusions are drawn in the last Section \ref{sec:conc}.

\section{The hard-rods model}
\label{sec:model}

Hard rods are the 1D counterpart of 3D hard spheres \cite{Mazzanti2008b,Mazzanti2008a}. 
The interparticle hard-rod potential is
\begin{equation}
V_{HR}(r) = \left\{
\begin{array}{cc}
\infty \quad &|r| \leq a \\
0      \quad &|r| >    a \\
\end{array}
\right.
\quad ,
\end{equation}
where $a$ is the rod size. 
The Hamiltonian of a system of $N$ particles inside an interval $[0,L]$ of length $L$ with 
interparticle HR potential is
\begin{equation}
\label{eq:hr_ham}
H = - \frac{\hbar^2}{2m} \, \sum_{i=1}^N \frac{\partial^2}{\partial r_i^2}
+ \sum_{i<j=1}^N V_{HR}(r_i-r_j) \quad , 
\end{equation}
where $m$ is the mass of the particles, and $(r_1 \dots r_N )$ $\in \mathbb{R}^N$ their coordinates.
The domain of the Hamiltonian operator \eqref{eq:hr_ham} is the set of wavefunctions $\Psi(r_1 \dots r_N) \in \mathcal{L}^2(\mathbb{R}^N)$ 
such that
\begin{equation}
\label{eq:hr_domain}
\begin{split}
&\Psi(r_1 \dots r_i \dots r_j   \dots r_N) = \pm \, \Psi(r_1 \dots r_j \dots r_i \dots r_N) \, , \\
&\Psi(r_1 \dots r_i \, + L \,\, \dots r_N) = \phantom{\pm} \, \Psi(r_1 \dots r_i \dots r_N) \, , \\
&\Psi(r_1 \dots r_i \dots r_j   \dots r_N) = 0 \,\,\,\, \mbox{if $|r_i-r_j| \leq a$} \, ,\\
\end{split}
\end{equation}
for any $i\neq j$.
The first of the conditions \eqref{eq:hr_domain} imposes Bose or Fermi symmetry, the second imposes periodic boundary conditions (PBC) and the third guarantees that $\langle \Psi | H | 
\Psi \rangle < \infty$.
Thanks to the second equation in \eqref{eq:hr_domain}, 
we can concentrate on positions $(r_1 \dots r_N ) \in \mathcal{C} =[0,L]^N$.

\subsection{Solution by Bethe Ansatz}
\label{subsec:bethe}

The solution of the HR Hamiltonian \eqref{eq:hr_ham} was first addressed by Nagamiya 
\cite{Nagamiya1940}, relying on the Bethe Ansatz method \cite{Bethe1931}. 
The author substituted PBC \eqref{eq:hr_domain} with slightly different boundary conditions, 
motivated by the study of particles arranged on a circle \cite{Nagamiya1940}.
The solution of the HR Hamiltonian by Bethe Ansatz was subsequently addressed by 
Sutherland in \cite{Sutherland1971b}, applying PBC.

In the present Section, we provide a detailed review of the solution of the HR model 
following the method of Refs.\cite{LiebLiniger1963,Lieb1963}, and a detailed description of its
elementary excitations. 
This is a key ingredient that permits to characterize the singularities of $S(q,\omega)$ predicted by the
nonlinear Luttinger liquid theory (Sec. \ref{sec:LLT}).

In order to solve the HR Hamiltonian \eqref{eq:hr_ham}, following Ref.\cite{Nagamiya1940}, let 
us concentrate on the sector $\mathcal{S}$ of the configuration space  $\mathcal{C}$ where
\begin{equation}
\begin{split}
&0         < r_1 < r_2-a \\
&r_{i-1}+a < r_i < r_{i+1}-a \quad i=2 \dots N-1 \\
&r_{N-1}+a < r_N < L-a \, ,\\
\end{split}
\end{equation}
which is related to all other sectors of the configuration space by a combination of
permutations and translations of the particles,
and eliminate the rod size $a$ by the transformation
\begin{equation}\label{eq:coordx}
x_i = r_i - (i-1) a \quad .
\end{equation}
The rod coordinates $x_i$ lie in the set
\begin{equation}
0 < x_1 < x_2 < \dots < x_N < L' \quad,
\end{equation}
where $L' = L-Na$ is called the unexcluded volume.
The HR Hamiltonian \eqref{eq:hr_ham} then takes the form
\begin{equation}
\label{eq:hr_ham2}
H = - \frac{\hbar^2}{2m} \, \sum_{i=1}^N \frac{\partial^2}{\partial x_i^2} \quad ,
\end{equation}
and the third condition \eqref{eq:hr_domain}, imposing that particles collide with each 
other as impenetrable elastic rods, can be correspondingly expressed as
\begin{equation}
\label{eq:hr_domain2}
\tilde{\Psi}(x_1 \dots x_i \dots x_j \dots x_N) = 0 \quad \mbox{if $x_i = x_j$} \quad ,
\end{equation}
where we introduce the notation
\begin{equation}
\Psi(r_1 \, r_2 \dots r_N) = \tilde{\Psi}(x_1,  \dots x_N), \quad (r_1 \, r_2 \dots r_N) \in \mathcal{S}
\end{equation}
to express $\Psi$ in terms of the rod coordinates. 
Eigenfunctions of \eqref{eq:hr_ham2} satisfying the condition \eqref{eq:hr_domain2} have 
the form \cite{Nagamiya1940}
\begin{equation}
\label{eq:hr_psiNAGprelim}
\tilde{\Psi}(x_1 \dots x_N) = 
\frac{1}{\sqrt{N!}} \, \mbox{det}\left( \frac{e^{i k_i x_j}}{\sqrt{L'}} \right)
\quad ,
\end{equation}
where $k_1 \dots k_N$ are a set of quantum numbers called quasi-wavevectors, that
will be identified later. 
The energy eigenvalue corresponding to \eqref{eq:hr_psiNAGprelim} is ${\energycommand_{\{k\}}=\frac{\hbar^2}{2m} 
\, \sum_{i=1}^N k_i^2}$. 
Moreover, \eqref{eq:hr_psiNAGprelim} is identically zero if and only if any two 
quasi-wavevectors coincide. 
The values of the quasi-wavevectors $k_1 \dots k_N$ are fixed imposing PBC to the 
wavefunctions \eqref{eq:hr_psiNAGprelim}. 
Practically, imposing PBC means requiring that
\begin{equation}
\label{eq:hr_PBC1}
\Psi(0 \, r_2 \dots r_N) = \Psi(L \, r_2 \dots r_N)
\end{equation}
for all $r_2 \dots r_N$.
Merging \eqref{eq:hr_psiNAGprelim} and \eqref{eq:hr_PBC1} one finds that PBC are satisfied 
if, for all quasi-wavevectors $k_i$, the following condition holds \cite{Sutherland1971b}
\begin{equation}
\label{eq:hr_PBC2}
(k_i - K) \, a = k_i (L-(N-1)a) - 2 \pi n_i + \xi^{B,F}(N) \quad ,
\end{equation}
where $i=1 \dots N$, $K = \sum_{i=1}^N k_i$, $n_i \in \mathbb{Z}$ is an integer number, 
$\xi^F(N) = 0$ and
\begin{equation}
\xi^B(N) = \left\{
\begin{array}{cc}
0   &\mbox{for $N$ odd\,\,} \\
\pi &\mbox{for $N$ even} \\
\end{array}
\right.
\quad .
\end{equation}
Equation \eqref{eq:hr_PBC2} leads easily to
\begin{equation}
\label{eq:hr_kPBC}
k_i = \frac{2 \pi}{L'} \, n_i - \frac{1}{L'} \, \xi^{B,F}(N) - \frac{aK}{L'} \quad .
\end{equation}
Remarkably, even if the quasi-wavevectors $k_i$ are constructed with both $L$ and $L'$, the total 
momentum $K$ is an integer multiple
\begin{equation}\label{eq:Ktot}
K = \frac{2 \pi}{L} \, \sum_{i=1}^N n_i - \frac{N \xi^{B,F}(N)}{L}
\end{equation}
of $\frac{2 \pi}{L}$.
To summarize, the eigenfunctions of the HR Hamiltonian are in one-to-one correspondence with combinations of $N$ integer numbers without repetition.

\subsection{Ground-state properties}

\label{ref:HR_GS}

For a system of $N$ Bose hard rods, the ground-state wavefunction 
is characterized by quasi-wavevectors
\begin{equation}
\label{eq:psiGS_qmom}
k_{i,GS} = \frac{2\pi}{L'} \, n_{i,GS} \quad\quad n_{i,GS} = - n_F + (i-1)
\end{equation}
symmetrically distributed around $0$, with ${n_F = (N-1)/2}$. The ground-state wavevector is naturally $K=0$. 
The ground-state energy reads \cite{Nagamiya1940,Mazzanti2008b}:
\begin{equation}
\label{eq:eos}
\begin{split}
\energycommand_{GS} &= 
\frac{\hbar^2}{2m} \left( \frac{2\pi}{L'} \right)^2 \frac{n_F (n_F+1) (2n_F+1)}{3}
\end{split}
\end{equation}

In the thermodynamic limit of large system size $N$ at constant linear density $\rho=N/L$, the ground-state energy per particle converges to
\begin{equation}
\label{eq:eos2}
\begin{split}
\energycommand_\infty = \lim_{N \to \infty} \frac{\energycommand_{GS}}{N} = 
\frac{\hbar^2 k_F^2}{6m (1-\rho a)^2} \quad ,
\end{split}
\end{equation}
where $k_F=\pi \rho$ is defined in analogy with the fermionic case.
The reduced dimensionality is responsible for the {\em{fermionization}} of impenetrable Bose particles: 
the strong repulsion between particles mimics the Pauli exclusion principle \cite{Girardeau1960}. In particular, the limit $\rho a = 0$ corresponds to the well-known Tonks-Girardeau gas, namely the hard-core limit of the Lieb-Liniger model \cite{LiebLiniger1963}, where all local properties are the same as for the ideal Fermi gas. At finite $\rho a$, we can think of HRs as evolving from the Tonks-Girardeau gas, in that the
infinitely strong repulsive interaction is accompanied by an increasing volume exclusion. HRs are therefore a model for the \emph{super} Tonks-Girardeau gas, which has been predicted and observed \cite{Astrakharchik2005,Batchelor2005,Tempfli2008,Haller2009,Panfil2013} as a highly excited and little compressible
state of the attractive Lieb-Liniger Bose gas, in which no bound states are present.

In the case of hard rods, the eigenfunctions of both Bose and Fermi systems have
the same functional form in the sector $\mathcal{S}$ of the configuration space;
away from $\mathcal{S}$, they differ from each other only by a sign associated to a
permutation of the particles \cite{Girardeau1960}.
Therefore, the matrix elements of local operators like the density fluctuation operator
\begin{equation}
{\rho}_q = \sum_{i=1}^N e^{-i q {r}_i}
\end{equation}
are identical for Bose and Fermi particles. This, in particular, implies that the dynamical structure factor
\begin{equation}
S(q,\omega) = \int_{-\infty}^\infty dt \, \frac{e^{i \omega t}}{2\pi N} \, \langle \Psi | e^{itH/\hbar}{\rho}_q e^{-itH/\hbar}{\rho}_{-q} | \Psi \rangle
\end{equation}
and the static structure factor
\begin{equation}
S(q) = \int_0^\infty d\omega S(q,\omega) = 
\frac{1}{N} \, \langle \Psi | {\rho}_{q} {\rho}_{-q} | \Psi \rangle
\end{equation}
are independent of the statistics.
Quite usefully for the purpose of QMC simulations in configuration space, the unnormalized bosonic ground-state wavefunction can be written in a Jastrow form for any $a$ \cite{Girardeau1960,Krotscheck1999,Mazzanti2008a}:
\begin{equation}\label{eq:exactjastrow}
 \Psi_{GS}(r_1\cdots r_N)= \prod_{i<j}|\sin{\pi(x_j-x_i)/L}|\quad.
\end{equation}

\subsection{Elementary excitations}\label{subsec:excit}

\begin{figure}
\begin{tikzpicture}[scale=0.7]
\shade [ball color=lightgray!30,opacity=0.5] ( 3,0) circle [radius=0.25];
\shade [ball color=lightgray!30,opacity=0.5] ( 1,0-2) circle [radius=0.25];
\shade [ball color=lightgray!30,opacity=0.5] (-3,0-4) circle [radius=0.25];
\draw[->,thick] (-4,0) -- (7,0);
\draw[->,thick] (-4,0-2) -- (7,0-2);
\draw[->,thick] (-4,0-4) -- (7,0-4);
\draw[black,dotted] ( 0,1) -- ( 0,-5);
\draw[black,dotted] (-3,1) -- (-3,-5);
\draw[black,dotted] ( 3,1) -- ( 3,-5);
\node at (7.5,-2) {$k$};
\shade [ball color=lightgray] (-3,0) circle [radius=0.25];
\shade [ball color=lightgray] (-2,0) circle [radius=0.25];
\shade [ball color=lightgray] (-1,0) circle [radius=0.25];
\shade [ball color=lightgray] ( 0,0) circle [radius=0.25];
\shade [ball color=lightgray] ( 1,0) circle [radius=0.25];
\shade [ball color=lightgray] ( 2,0) circle [radius=0.25];
\shade [ball color=lightgray] ( 6,0) circle [radius=0.25];
\draw[->,thick]  (3,0.35) to [out=45,in=135] (6,0.35);
\node at (-3,-5.5) {$-k_F$};
\node at ( 0,-5.5) {$ 0  $};
\node at ( 3,-5.5) {$ k_F$};
\shade [ball color=lightgray] (-3,0-2) circle [radius=0.25];
\shade [ball color=lightgray] (-2,0-2) circle [radius=0.25];
\shade [ball color=lightgray] (-1,0-2) circle [radius=0.25];
\shade [ball color=lightgray] ( 0,0-2) circle [radius=0.25];
\shade [ball color=lightgray] ( 3,0-2) circle [radius=0.25];
\shade [ball color=lightgray] ( 2,0-2) circle [radius=0.25];
\shade [ball color=lightgray] ( 4,0-2) circle [radius=0.25];
\draw[->,thick]  (1,0.35-2) to [out=45,in=135] (4,0.35-2);
\shade [ball color=lightgray] ( 1,0-4) circle [radius=0.25];
\shade [ball color=lightgray] (-2,0-4) circle [radius=0.25];
\shade [ball color=lightgray] (-1,0-4) circle [radius=0.25];
\shade [ball color=lightgray] ( 0,0-4) circle [radius=0.25];
\shade [ball color=lightgray] ( 3,0-4) circle [radius=0.25];
\shade [ball color=lightgray] ( 2,0-4) circle [radius=0.25];
\shade [ball color=lightgray] ( 4,0-4) circle [radius=0.25];
\draw[->,thick]  (-3,0.35-4) to [out=22.5,in=157.5] (4,0.35-4);
\end{tikzpicture}
\caption
{
Pictorial representation of the Lieb-$I$ excitation with $p = 6$ (top) and of the Lieb-$II$ 
excitation (middle) with $h=5$ for $N=7$ HRs.
The Lieb-$II$ excitation with $h=1$ is called umklapp excitation (bottom).
} \label{fig:exc}
\end{figure}
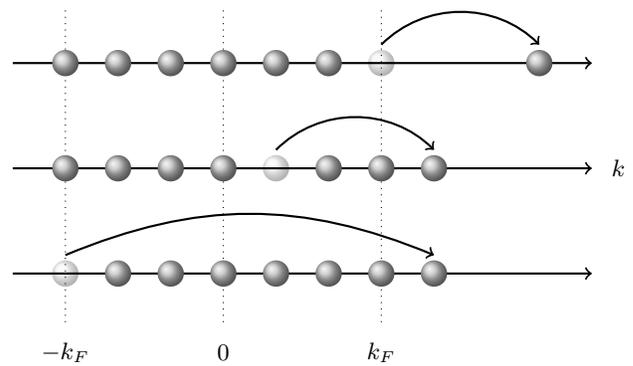

In the previous Subsection \ref{ref:HR_GS}, we have recalled that the ground-state wavefunction 
of the HR system, once expressed in terms of the rod coordinates, has the functional form of 
a Fermi sea with renormalized coordinates and wavevectors $k_{i,GS}$, specified in terms 
of integer numbers $n_{i,GS}$.
The excited states of the system are obtained creating single or multiple particle-hole pairs
on top of this pseudo Fermi sea \cite{CastroNeto1994}.
The simplest excitations, illustrated in Figure \ref{fig:exc}, consist in the creation of 
a single particle-hole pair
\begin{equation}
\label{eq:phpair1}
n_{i,PH} = n_{i,GS} + (p - n_{h,GS} ) \delta_{i,h} \quad ,
\end{equation}
 where $h \in \{ 1 \dots N \}$ is the index of the original quantum number to be modified (``hole'') and $p > n_F$ is the new integer quantum number of index $h$ (``particle''). Among single particle-hole excitations, a role of great importance in the interpretation of 
$S(q,\omega)$ is played by the following {\emph {Lieb-I}}
\begin{equation}
\label{eq:phpair2}
n_{i,I} = n_{i,GS} + (p - n_F) \, \delta_{i,N}
\end{equation}
and {\emph {Lieb-II}}
\begin{equation}
\label{eq:phpair3}
n_{i,II} = n_{i,GS} + (n_F+1 - n_{h,GS}) \, \delta_{i,h}
\end{equation}
modes, which are reminiscent of the corresponding excitations of the Lieb-Liniger
model \cite{Lieb1963}.
As shown in Figure \ref{fig:exc}, in the Lieb-I excitation, a rod is taken from the 
Fermi level $h=N$ to some high-energy state associated to an integer $p > n_F$, while 
in the Lieb-II excitation a rod is taken from a low-energy state $h_{II}$ to just 
above the Fermi level $p=n_F+1$.
In both cases, in view of the collective nature of the quasi-wavevectors $k_i$, 
the excitation of that rod provokes a recoil of all the other rods, according to Equations
\eqref{eq:hr_kPBC} and \eqref{eq:Ktot}.

A simple calculation shows that the dispersion relation of the Lieb-I excitation
is given by 
\begin{equation}
\energycommand^{(N)}_I(q) = \energycommand_I(q) + \Delta \energycommand^{(N)}_I(q)
\end{equation}
where $q \geq 0$ is the wavevector of the excitation. The dispersion relation 
$\energycommand_I(q)$ in the thermodynamic limit reads
\begin{equation}
\frac{\energycommand_I(q)}{\energycommand_F} = \frac{4}{K_L} \,
\left( x + x^2 \right)
\quad ,
\end{equation}
In the previous equation $\energycommand_F = \frac{\hbar^2 k_F^2}{2m}$, $x = \frac{q}{2k_F}$ and the physical
meaning of the Luttinger parameter $K_L = (1-\rho a)^2$ will be elucidated in Section \ref{sec:LLT}.
Size effects $\Delta \energycommand^{(N)}_I(q)$ have the form
\begin{equation}
\frac{\Delta \energycommand^{(N)}_I(q)}{\energycommand_F} = 
- \frac{2}{N K_L} \, x \, \left( 1 + (1-K_L) x \right)
\end{equation}
whenever $\xi^{B,F}(N)=0$.
A similar calculation shows that the dispersion relation of the Lieb-II excitation
is given by $\energycommand^{(N)}_{II}(q) = \energycommand_{II}(q) + \Delta \energycommand^{(N)}_{II}(q)$
with $0 \leq q \leq 2 k_F$ and
\begin{equation}
\begin{split}
\frac{\energycommand_{II}(q)}{\energycommand_F} &=
\frac{4}{K_L} \, \left( x - x^2 \right)
\quad , \\
\frac{\Delta \energycommand_{II}(q)}{\energycommand_F} &=
\frac{ \energycommand_{II}(q) }{N}  + \frac{x^2}{N}
\end{split}\label{eq:liebIIfinite}
\end{equation}

Other relevant excitations are those producing {{supercurrent states}}
\cite{Lieb1963,Vignale2005,Cherny2011}
\begin{equation}
n_{i,SC} = n_{i,GS} + s \quad\quad s \in \mathbb{N}
\end{equation}
with momenta $q = 2 s k_F$ and excitation energies $\energycommand_{sc}(q) = 
\frac{\hbar^2}{2m} \frac{q^2}{N}$ {{independent of the rod length $a$}}, and
vanishing in the thermodynamic limit.
Supercurrent states correspond to Galilean transformations of the ground state
with velocities $v_{SC} = \frac{2 \hbar k_F}{m} \, \frac{s}{N}$.
The first supercurrent state, in particular, is also termed {{umklapp excitation}}
\cite{Lieb1963,Vignale2005,Cherny2011}.

In the thermodynamic limit, the particle-hole 
excitations \eqref{eq:phpair1} span the region $\omega^*_-(q) \leq \omega \leq 
\omega^*_+(q)$ of the $(q,\omega)$ plane, where 
\begin{equation}\label{eq:particleholeHR}
\frac{\hbar\omega_\pm^*(q)}{\energycommand_F} =
\frac{4}{K_L} \, \left| \frac{q}{2k_F} \pm \left(\frac{q}{2k_F}\right)^2 \right| \quad.
\end{equation}
The curves $\hbar\omega_\pm^*(q)$ have the same functional form of the ideal Fermi gas 
particle-hole boundaries $\hbar\omega_\pm(q)=|\hbar^2 k_F q/m\pm\hbar^2q^2/2m|$, except 
for the substitution of the bare mass $m$ with $m^* = m K_L < m$, as we argued in Ref.~\cite{Bertaina2016}.

The upper branch of this renormalized particle-hole continuum coincides with the Lieb-I
mode. For $q\leq 2 k_F$, its lower branch coincides with the Lieb-II mode and, for 
$q \geq 2 k_F$, with the particle-hole excitations
\begin{equation}
n_{i,PH} = n_{i,GS} + 1 + (p - n_{N,GS} - 1) \delta_{i,N} \quad ,
\end{equation}
resulting from the combination of the Lieb-I and the umklapp modes.

It is worth noticing that the Lieb-II dispersion relation constitutes the energy threshold for excitations for $0<q<2k_F$. Away from this 
basic region, the energy threshold in the thermodynamic limit can be obtained by a combination of inversions and shifts \cite{Imambekov2009}, 
and corresponds to a combination of a Lieb-II mode and multiple umklapp excitations. To summarize, the low-energy threshold is given by
\begin{equation}
\label{eq:lowenthr}
\frac{\hbar \omega_{th}(q)}{\energycommand_F} = \frac{4}{K_L}\left(\frac{q^*_n}{2k_F} - 
\left( \frac{q^*_n}{2k_F} \right)^2 \right)
\quad ,
\end{equation}
where $2 n k_F \leq q \leq 2 (n+1) k_F$ and $q_n^* = q - 2 n k_F$. Finite size corrections are the same as in Eq.~\eqref{eq:liebIIfinite} \cite{Bertaina2016}.

Remarkably, \eqref{eq:lowenthr} corresponds also to the dispersion relation of dark solitons of 
composite bosons in Yang-Gaudin gases of attractively interacting fermions in the deep molecular regime, 
even though in that case the molecular scattering length is negative (corresponding to a repulsive Lieb-Liniger molecular gas)
\cite{Brand2016}.

\subsection{Comparison with Luttinger liquid theories}

\label{sec:LLT}

The low-energy excitations of a broad class of interacting 1D systems are captured 
by the phenomenological TLL field theory \cite{Tomonaga1950,Luttinger1963,Mattis1965,Haldane1983,Haldane1983b,Imambekov2012}.
The TLL provides a universal description of interacting Fermi and Bose particles 
by introducing 
two fields, ${\phi}(x)$ and ${\theta}(x)$ representing the density and phase 
oscillations of the destruction operator ${\Psi}(x) \simeq \sqrt{\rho + \partial_x{\phi}(x)} e^{i {\theta}(x)}$,
and a quadratic low-energy Hamiltonian describing the dynamics of those fields
\begin{equation}
\label{eq:lutham}
{H}_{LL} = \frac{\hbar}{2\pi} \int dx \left(  c K_L \partial_x{\theta}(x)^2 + \frac{c}{K_L} \partial_x{\phi}(x)^2 \right)
\quad .
\end{equation}
For Galilean-invariant systems, the sound velocity $c$ is related to the positive Luttinger parameter $K_L$ through $c = \frac{v_F}{K_L}$. The quadratic nature of \eqref{eq:lutham} allows for the calculation of correlation 
functions and thermodynamic properties in terms of $c$ and $K_L$.
Within the TLL theory, in the low-momentum and low-energy regime $S(q,\omega)$ features
collective phonon-like excitations $\omega_{LL}(q) = c |q|$ with sound velocity $c$. 

The TLL theory has been recently extended \cite{Imambekov2009,Imambekov2012}
beyond the low-energy limit, where the assumption of linear excitation 
spectrum $\omega_{LL}(q)$ is not sufficient for accurately predicting dynamic response functions.
Assuming that,
for any momentum $q$, $S(q,\omega)$ has support above a low-energy 
threshold $\omega_{th}(q)$ and
interpreting excitations with momentum $q$ between $2n k_F$ and $2nk_F + 2 k_F$
as the creation of mobile holes of momentum $q_n^* = q - 2 n k_F$ coupled with the 
TLL \cite{Imambekov2009,Imambekov2012}, it is possible to show that for a broad class of 
Galilean-invariant systems $S(q,\omega)$ features a power-law singularity close to the low-energy 
threshold $\omega_{th}(q)$ with the following functional form:
\begin{equation}\label{eq:powerlaw}
S(q,\omega) = \theta\left(\omega-\omega_{th}(q_n^*)\right) \, 
\left| \omega - \omega_{th}(q_n^*) \right|^{-\mu_n(q)} \quad ,
\end{equation}
where the exponent
\begin{equation}
\label{eq:nllt1}
\begin{split}
\mu_n(q) = 1 & - 
   \frac{1}{2} \left(   (2n+1) \sqrt{K_L}  + \frac{\delta_+(q_n^*) + \delta_-(q_n^*)}{2\pi} \right)^2 \\
&- \frac{1}{2} \left( \frac{1}{\sqrt{K_L}} + \frac{\delta_+(q_n^*) - \delta_-(q_n^*)}{2\pi} \right) \\
\end{split}
\end{equation}
is specified in terms of the phase shifts
\begin{equation}
\label{eq:nllt2}
\frac{\delta_\pm(q)}{2\pi} = 
\frac{\frac{1}{\sqrt{K_L}} \left( \frac{\hbar q}{m} + \frac{\partial \omega_{th}(q)}{\partial q} \right)
\pm \sqrt{K_L} 
\left( \frac{v_s}{K_L}  - \frac{1}{\pi} \frac{\partial \omega_{th}(q)}{\partial \rho} \right)}
{2 \left( \mp \frac{\partial \omega_{th}(q)}{\partial q} - v_s \right) }
\end{equation}
The only phenomenological inputs required by the nonlinear TLL theory are the Luttinger parameter $K_L$ 
and the low-energy threshold $\omega_{th}(q)$, which in the case of hard rods are exactly known. 
Namely, we recall that the Luttinger parameter $K_L$ \cite{Haldane1983,Haldane1983b} can be computed from the compressibility
\begin{equation}
\kappa_S^{-1} = \rho \, \frac{\partial}{\partial\rho}\left( \rho^2 \, 
\frac{\partial \energycommand_\infty}{\partial\rho} \right)
\end{equation}
through the formula $m K_L^2 = \hbar^2 \, k_F^2 \, \rho \, \kappa_S$. The resulting exact expression, 
$K_L = (1-\rho a)^2$ \cite{Mazzanti2008a}, provides a Luttinger parameter always smaller than $1$, and converging towards 
$0$ as the excluded volume $Na$ converges towards $L$.
Notice that both Lieb-I and Lieb-II dispersions approach $q = 0$ with slope equal to the sound velocity 
$c= \frac{v_F}{K_L} > v_F$. The low-energy threshold, Eq. \eqref{eq:lowenthr}, has been described in the previous Section.

\begin{figure}
\centering
\includegraphics[width=0.50\textwidth]{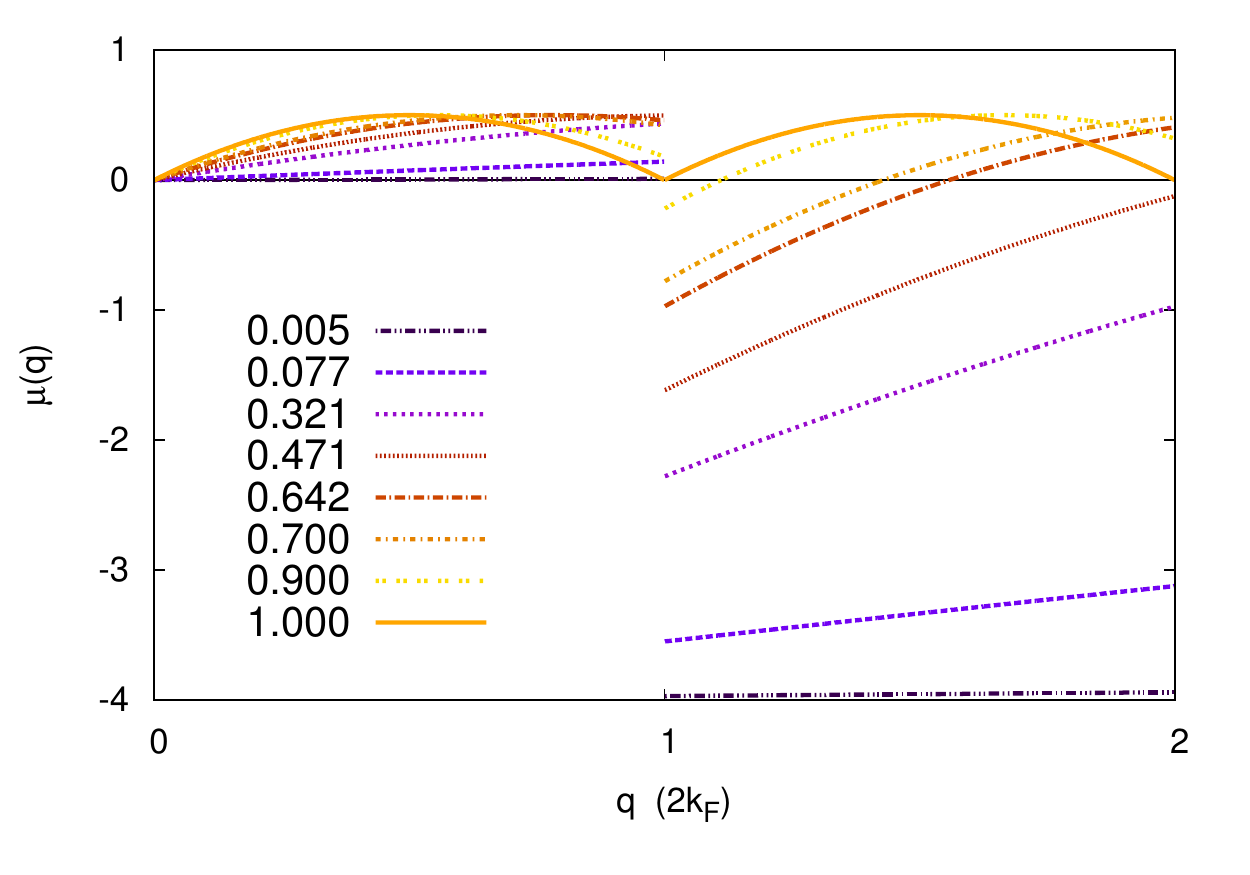}
\caption{
(color online) Nonlinear Luttinger theory exponents $\mu(q)$ for hard rods at the studied
densities from $\rho a = 0.077$ (blue solid line) to $\rho a = 0.900$ (red solid line).
Momenta are measured in units of $2 k_F = 2\pi \rho$.
The $\rho a \to 1$ limit is also shown (solid line).
Notice that the densities $\rho a \geq 0.5$ have a special wavevector $\mathcal{Q}_1$
between $2k_F$ and $4k_F$.
}
\label{fig:mu_hr}
\end{figure}

Knowledge of these two quantities permits to compute the exponent 
$\mu_n(q)$ exactly from \eqref{eq:nllt1} and \eqref{eq:nllt2}. We find \cite{Bertaina2016}
\begin{equation}
\label{eq:mu_hr}
\mu_n(q) = - 2 \left( \tilde{q} - n \right) \, \left( \tilde{q}- (n+1) \right) \quad ,
\end{equation}
with $\tilde{q} = \frac{qa}{2\pi}$. In Fig. \ref{fig:mu_hr} we show the power-law exponents for momenta $0<q<4k_F$ and different densities. Notice that the functional form of \eqref{eq:mu_hr} is that of a sequence of parabola arcs, intersecting null values at the special momenta $\mathcal{Q}_n=n 2\pi/a$, with integer $n<\rho a/(1-\rho a)$. Such momenta, even for larger $n$, have already been recognized to be special \cite{Mazzanti2008a}, in that they admit the exact calculation of $S(\mathcal{Q}_n)$ and $S(\mathcal{Q}_n,\omega)$. Namely, for those special momenta, the HRs at density $\rho$ behave as an ideal Fermi gas with increased density $\rho^\prime=\rho/(1-\rho a)$.

It is worth pointing out that
TLL theories have limits of applicability, and thus do not exhaust our understanding of 1D substances 
\cite{Caux2006,Fabbri2015,Meinert2015}.
The investigation of dynamical properties like $S(q,\omega)$ beyond the limits of applicability of 
Luttinger liquid theories, where the Physics is non-universal, typically requires numerical calculations or 
QMC simulations \cite{Caux2006,DePalo2008,Bertaina2016}.

\section{Methods}
\label{sec:method}

In the present work, the zero-temperature dynamical structure factor of a system of Bose HRs is 
calculated using the {\em{exact}} Path Integral Ground State (PIGS) QMC method to compute 
imaginary-time correlation functions of the density fluctuation operator, and the 
state-of-the-art Genetic Inversion via Falsification of Theories (GIFT) analytic continuation 
method to extract the dynamical structure factor.

This approach, which we briefly review in this Section, has provided robust 
calculations of dynamical structure factors for several non-integrable systems like 1D, 2D 
and 3D He atoms \cite{Bertaina2016,Vitali2010,Nava2013,Arrigoni2013} and hard spheres \cite{Rota2013}.

The PIGS method is a projection technique in imaginary time that, starting from 
a trial wavefunction $\Psi_T(R)$, where $R=(r_1 \dots r_N ) \in \mathcal{C}$
denotes a set of spatial coordinates of the $N$ particles, projects it onto the
ground-state wavefunction $\Psi_{GS}(R)$ after evolution over a sufficiently
long imaginary-time interval $\tau$ \cite{Sarsa2000,Galli2003,Patate}.
In typical situations, the functional form of $\Psi_T(R)$ is guessed combining
physical intuition and mathematical arguments based on the theory of stochastic
processes \cite{Holzmann2003}. $\Psi_T(R)$ is then specified by
one or more free parameters, that are chosen using suitable optimization algorithms
\cite{Kalos1986,Toulouse2007,Motta2015}.

In the case of HRs, knowledge of the exact ground-state wavefunction \eqref{eq:exactjastrow} makes the 
projection of a trial wavefunction $\Psi_T(R)$ approximating the ground state 
of the system unnecessary. 
However, the PIGS method can be used to give unbiased estimates of the 
density-density correlator
\begin{equation}
\label{eq:pigsfqt}
\begin{split}
F(q,\tau) &=
\langle \Psi_{GS} | e^{\tau H} {\rho}_{-q} \, e^{- \tau {H} } \, {\rho}_{q} | \Psi_{GS} \rangle = \\
&= \frac{ \int dR_M dR_0 \, p(R_M,R_0) \,  \rho_{-q}(R_M) \, \rho_q(R_0) }{ \int dR_M dR_0 \, p(R_M,R_0) } \quad ,
\end{split}
\end{equation}
with $p(R_M,R_0) = \Psi_{GS}(R_M) G(R_M,R_0;\tau) \Psi_{GS}(R_0)$
and $G(R_M,R_0;\tau) = \langle R_M | e^{-\tau {H} } | R_0 \rangle $. 

The propagator $G(R',R;\tau)$ is in general not known, but suitable approximate expressions are available for small $\delta \tau = \tau/M$, where $M$ is a large integer number.
Using one of these expressions in place of the exact propagator is the only approximation characterizing the calculations of the present work. The method is exact though, since this approximation affects the computed expectation values to 
an extent which is below their statistical uncertainty and such regime is always attainable by taking $\delta\tau$ sufficiently small. 
Then, the convolution formula permits to express $G(R_M,R_0;\tau)$ as
\begin{equation}
G(R_M,R_0;\tau) = \int dR_{M-1} \dots dR_1 \prod_{i=0}^{M-1}G(R_{i+1},R_i;\delta\tau) \, ,
\end{equation}
whence the PIGS estimator of $F(q,\tau)$ takes the form
\begin{equation}
\label{eq:pigsfqtp}
\begin{split}
F(q,\tau) = \frac{ \int d X \, p(X) \, \rho_{-q}(R_M) \, \rho_q(R_0) }{ \int dX \, p(X) } \, .
\end{split}
\end{equation}
In \eqref{eq:pigsfqtp},
$X = (R_0 \dots R_M)$ denotes a path in the configuration space $\mathcal{C}$ of the
system, and
\begin{equation}
p(X) = \Psi_{GS}(R_M) \prod_{i=0}^{M-1}G(R_{i+1},R_i;\delta\tau) \Psi_{GS}(R_0)
\end{equation}
can be efficiently sampled using the Metropolis
algorithm \cite{Metropolis1953}.
In the present work, we have employed the pair-product approximation \cite{PP} 
to express the propagator relative to a small time step $\delta\tau$ as
\begin{equation}
\label{eq:ppprop}
G(R,R';\delta\tau) =
\prod_{i=1}^N G_0(r_i,r'_i;\delta\tau) 
\prod_{i<j}^N G_{\rm rel}(r_{ij},r'_{ij};\delta\tau)
\quad ,
\end{equation}
where $G_0$ is the free-particle propagator
\begin{equation}
 G_0(r,r';\delta\tau) = \frac{1}{\sqrt{2\pi\lambda\delta\tau}}
e^{-(r-r')^2/4\lambda\delta\tau}
\end{equation}
with $\lambda= \hbar^2/2m$, and $G_{\rm rel}$ is obtained from the exactly known 
solution of the two-body scattering problem, similarly to a standard approach for hard spheres in 3D \cite{CaoBerne1992}
\begin{equation}
G_{\rm rel}(r,r';\delta\tau) = 
1 - e^{-\frac{(r-a)(r'-a)}{2\lambda\delta\tau}} \quad.
\end{equation} 
Moreover, in order to select an appropriately small $\delta\tau$, we have both checked the convergence of the static structure factor and the convergence of energy when the exact initial trial wavefunction is replaced with an approximate one \cite{Gordillo2012}.

The initial imaginary-time value of Eq. \eqref{eq:pigsfqt} is the static 
structure factor $F(q,0)=S(q)$. For finite values of $\tau$, $F(q,\tau)$ is instead related 
to $S(q,\omega)$ by the Laplace transform
\begin{equation}
\label{eq:invlap}
F(q,\tau) = \int_0^\infty d\omega \, e^{-\tau \omega} \, S(q,\omega) \quad .
\end{equation}
Equation \eqref{eq:invlap} should be inverted in order to determine $S(q,\omega)$
from $F(q,\tau)$. However, it is well-known that such inverse problem is {\em{ill-posed}}, 
in the sense that many different trial dynamical structure factors, ranging from featureless 
to rich-in-structure distributions, have a forward Laplace transform which is compatible 
with the QMC results for $F(q,\tau)$: there is not enough information to find a unique 
solution of \eqref{eq:invlap} \cite{Tarantola2006,MEM1,MEM2,MEM3,MEM4,Mishchenko2000,Vitali2010}.
Different methodologies have been used to extract real-frequency response functions from imaginary-time correlators; in the present work, we rely on the Genetic Inversion via Falsification of Theories (GIFT) method \cite{Vitali2010}. The aim of GIFT is to 
collect a large collection of such dynamical structure factors in order to discern the presence 
of common features (e.g. support, peak positions, intensities and widths). The GIFT method has been applied to the study of liquid ${}^4 \mbox{He}$ \cite{Overpress,Anisotropic,Bertaina2016}, 3D hard spheres \cite{Rota2013},
2D Yukawa Bosons \cite{Molinelli2016}, liquid ${}^3 \mbox{He}$ \cite{Nava2013}, 2D soft disks \cite{Saccani2011,Saccani2012}, the 2D Hubbard model \cite{Vitali2016} and 1D soft rods \cite{Teruzzi2016}, in all cases providing very accurate reconstructions of $S(q,\omega)$ or the single particle spectral function. Recently \cite{Bertaina2016}, we have shown that in 1D, when $\omega_{th}(q)$ is known, also the shape close to the frequency threshold can be approximately inferred. This is the reason why in subsection \ref{subsec:excit} we insisted on the calculation of $\omega_{th}(q)$ for a finite system, which is a most useful quantity in our approach.  
Details of the GIFT method can be found in \cite{Vitali2010,Bertaina2016}. As in \cite{Bertaina2016}, we have used genetic operators which are able to better describe broad features typical of 1D systems. Moreover, the set of discrete frequencies of the model spectral functions used in the algorithm has been extended to non-equispaced frequencies, in order to better describe the regions where most of the weight accumulates. 

\begin{table}[bp]
\begin{tabular}{ll}
\hline\hline
$\rho a$   & $K_L$       \\
\hline
0.005 & 0.990 \\
0.077 & 0.852 \\
0.321 & 0.461 \\
0.471 & 0.280 \\
0.642 & 0.128 \\
0.700 & 0.090 \\
0.900 & 0.010 \\
\hline\hline
\end{tabular}
\caption{Densities and values of $K_L$ studied in the present work.}
\label{Table:1}
\end{table}

\section{Results}
\label{sec:results}

We computed $F(q,\tau)$ for systems of $N=50$ hard rods at densities $\rho a$ listed in Table \ref{Table:1},
and wavevectors $q \leq 8 k_F$. Before describing our results on the dynamical structure factor, we demonstrate the accuracy of our calculations 
by analyzing in detail finite-size effects on static properties which have already been studied in Ref. \cite{Mazzanti2008a}.

\begin{figure}[t]
\centering
\includegraphics[width=0.48\textwidth]{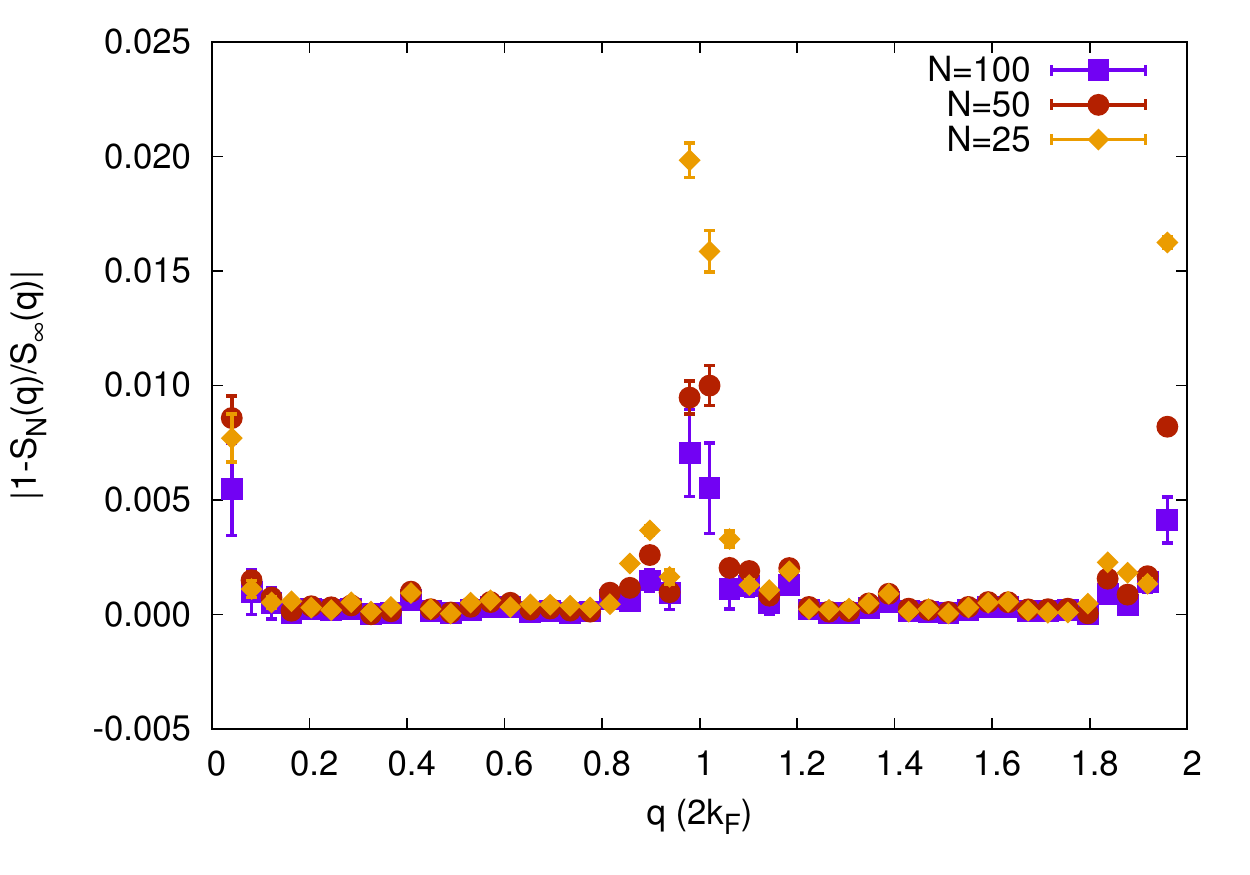}
\caption{
(color online) 
We quantify the finite-size effects on $S(q)$ at the representative density $\rho a = 0.700$
computing the relative error $|S_N(q) - S_\infty(q) | / S_\infty(q)$, where $S_\infty(q) = 
\lim_{N\to \infty} S_N(q)$ is extrapolated. 
Away from the points $q=2k_F,4k_F$, the relative error is below $1\%$ for $N=50$ particles.
}
\label{fig:scal}
\end{figure}

\subsection{Assessment of accuracy}

Results are affected by very weak finite-size effects, and thus are well representative 
of the thermodynamic limit.
For example, \eqref{eq:eos} and \eqref{eq:eos2} yield the following finite-size corrections
to the ground-state energy
\begin{equation}
\frac{\energycommand_{GS}}{N} = \energycommand_\infty \left( 1 - \frac{1}{N^2} \right)
\end{equation}
whence $\frac{\energycommand_{GS}}{50} = 0.9996 \, \energycommand_\infty$.
To further assess the finite-size effects on our results, in Figure \ref{fig:scal} we 
compute the static structure factor of $N =$ $50$, $100$ hard rods at $\rho a 
= 0.700$ using the VMC method, which is an exact method when the exact (ground state) wave function is known, as in this case. 

\begin{figure}[b]
\centering
\includegraphics[width=0.48\textwidth]{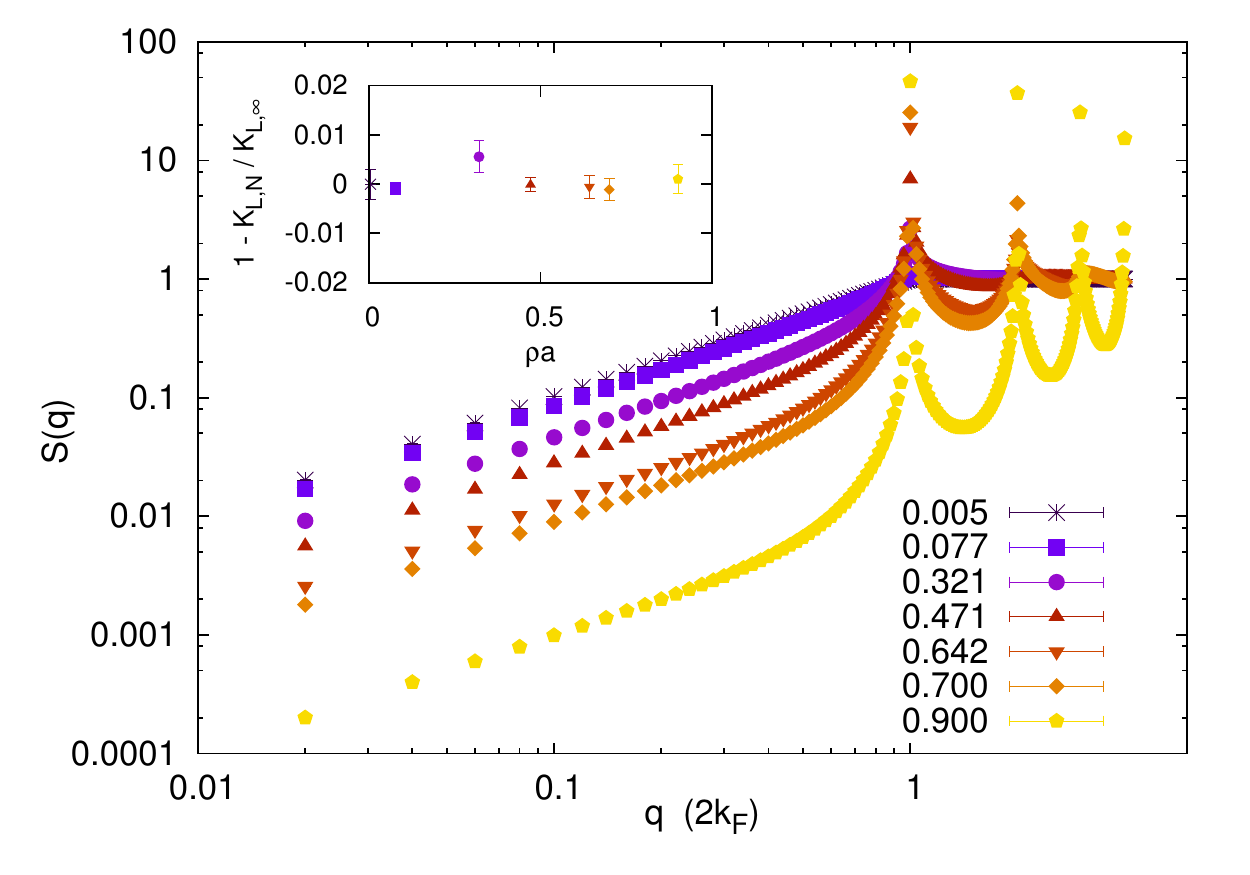}
\caption{
(color online)
Static structure factor of $N=50$ rods at $\rho a = 0.005$, $0.077$, $0.321$, $0.471$, 
$0.642$, $0.700$, $0.900$.
Inset: relative error on the Luttinger parameter, computed from the low-momentum behavior of 
the static structure factor $S(q) \simeq K_L q/(2k_F)$. 
In all cases, the relative error is below $1 \%$.
}
\label{fig:skx}
\end{figure}

At $q = 2 k_F, 4 k_F$, the static structure factor displays peaks of diverging weight as
predicted by the TLL theory \cite{Mazzanti2008a,Astrakharchik2014}:
\begin{equation}
\label{eq:skpeaks}
\begin{split}
S(2mk_F) &= S_{smooth}(2mk_F) + S_{peak}(2mk_F) = \\
         &= S_{smooth}(2mk_F) + C_m \, N^{1-2m^2 K_L} 
\end{split}
\end{equation}
Away from those points, the VMC estimates of the static structure factor are compatible
with each other and with the extrapolation of $S(q)$ to the thermodynamic limit, within 
the error bars of the simulations, reflecting the weakness of finite-size effects.
Moreover, in Figure \ref{fig:skx} we show that static structure factors of $N=50$ rods
permit to compute the Luttinger parameter $K_L$ without appreciable finite-size effects.

\begin{figure}
\centering
\includegraphics[width=0.5\textwidth]{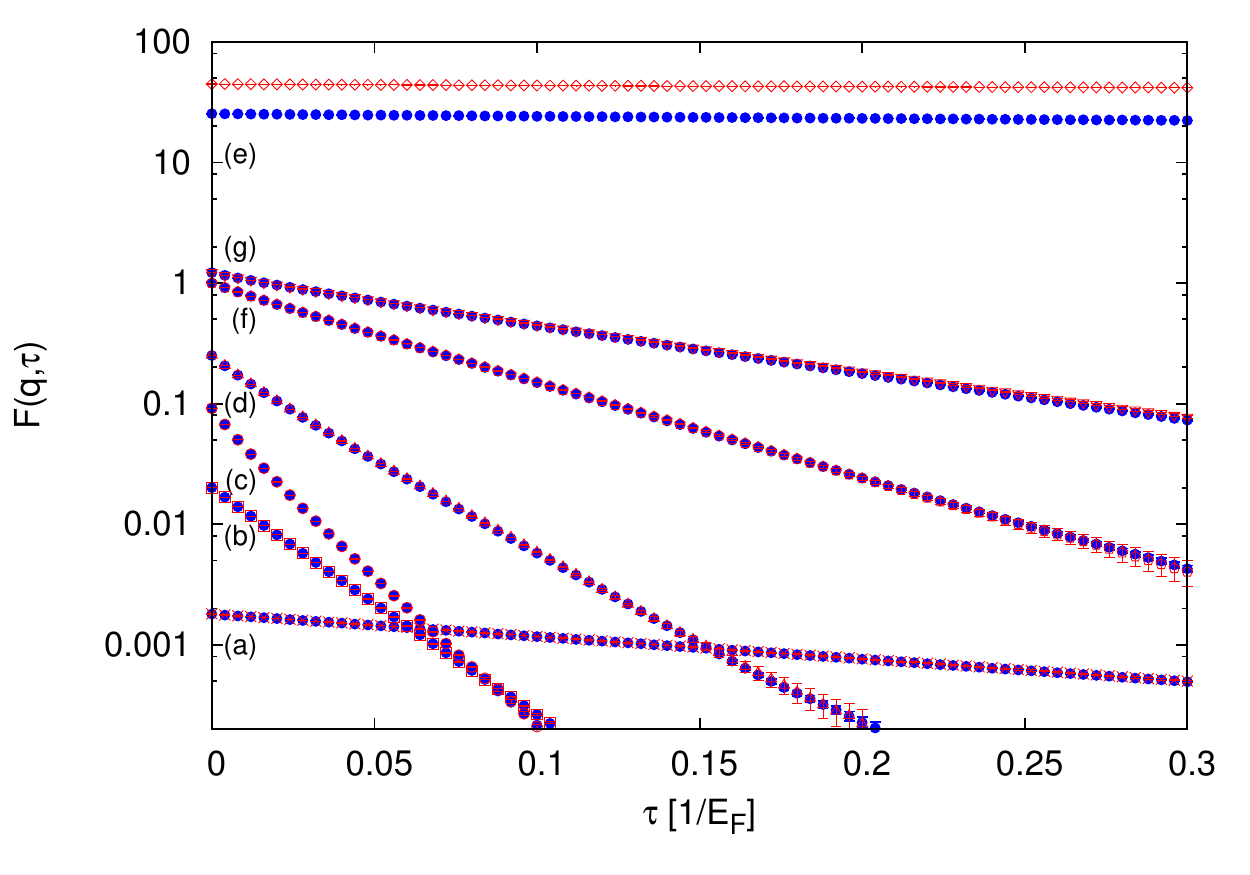}
\caption{
(color online)
$F(q,\tau)$ of $N=50,100$ (blue circles, red open diamonds) rods at $\rho a = 0.700$ for $\frac{q}{2k_F} = 
0.02$, $0.22$, $0.62$, $0.82$, $0.96$, $1.00$, $1.08$ (a to g). 
At $q = 2k_F$ we observe strong finite-size effects, originating from the quasi-Bragg 
peaks in $S(q)$.
Away from $q = 2k_F$, in the relevant imaginary-time interval $\tau \energycommand_F \leq 0.3$
the two $F(q,\tau)$ are in satisfactory agreement.
}
\label{fig:fqt}
\end{figure}

The same favorable behavior is exhibited by $F(q,\tau)$. In Figure \ref{fig:fqt}, we show
$F(q,\tau)$ for $N=50,100$ rods at the representative density $\rho a=0.700$. 
Away from $q = 2k_F$ the two systems have statistically compatible $F(q,\tau)$, confirming 
the modest entity of finite-size effects.

\subsection{Dynamical structure factors}\label{sec:spectra}

\begin{figure*}
\begin{center}
\includegraphics[width=0.86\textwidth]{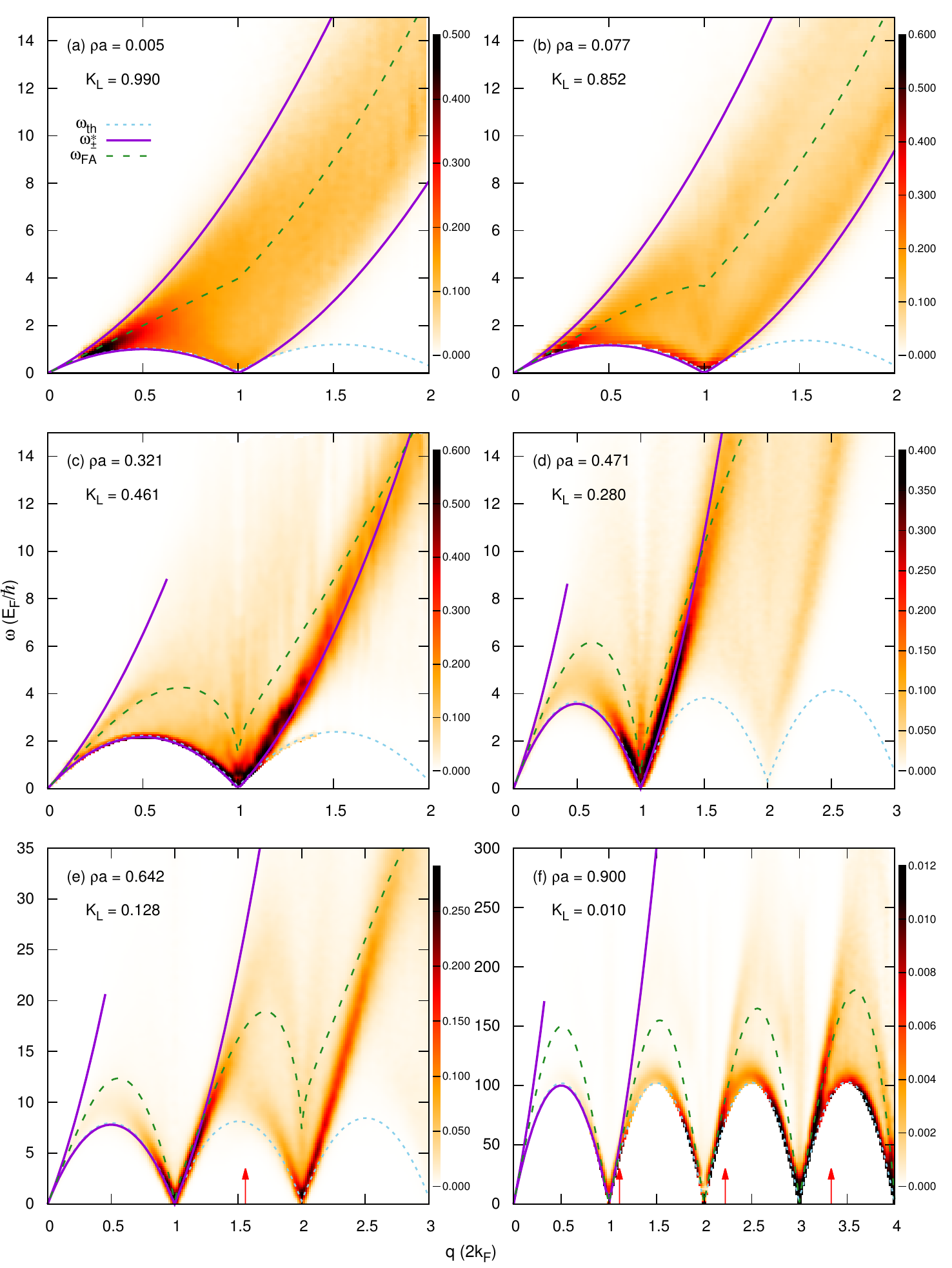}
\caption{(color online)
Color map of the dynamic structure factor, in units of $\hbar/E_F$, at $\rho a = 0.005$, $0.077$, $0.321$, $0.471$, $0.642$, $0.900$
(left to right, top to bottom). Momenta are measured in units of $2 k_F = 2\pi \rho$ and energies in units of $E_F$.
The low-energy threshold (blue dotted line), the branches $\omega_\pm^*(q)$ (purple solid lines) and the Feynman
approximation for the excitation spectrum (green dashed line) are drawn for comparison. Panels (e), (f) also 
show the special wavevectors $\mathcal{Q}_n$ (red arrows).}
\label{fig:spectra}
\end{center}
\end{figure*}

In Figure \ref{fig:spectra} we show the dynamical structure factor at densities ranging
from $\rho a = 0.005$ to $0.900$.
At all the studied densities, for momentum $q<2k_F$, $S(q,\omega)$ has most of the spectral weight inside the particle-hole band 
$\omega^*_-  (q) \leq \omega \leq \omega^*_+(q)$
spanned by single particle-hole excitations, Eq.~\eqref{eq:particleholeHR}. Contributions from multiple particle-hole excitations become relevant at high densities and momenta.

At the lowest density, panel (a), the spectral weight is broadly distributed inside the 
particle-hole band, showing a behavior reminiscent of the Tonks-Girardeau model of 
impenetrable point-like bosons, to which the HR model reduces in the $\rho a \to 0$ limit.

The low momentum and energy behavior can be understood in the light of the 
nonlinear TLL theory: as illustrated in Figure \ref{fig:mu_hr}, for momenta $q< 2k_F$ it 
predicts a power-law behavior for $S(q,\omega)$, with an exponent \eqref{eq:mu_hr} slightly 
larger than zero.
This prediction is consistent with the spectrum in panel (a), showing a weak concentration of spectral weight close to the low-energy threshold for $q<2k_F$.
For $q>2k_F$, the support of $S(q,\omega)$ departs from the low-energy threshold as 
it can be seen in panel (a), where the spectral weight remains concentrated inside the 
particle-hole band for all $q$.
Correspondingly, for $2k_F<q<4k_F$, the nonlinear TLL predicts a large negative exponent, suggesting
the absence of spectral weight in the proximity of the low-energy threshold.

A similar behavior is shown at density $\rho a = 0.077$, panel (b), where the spectral
weight concentrates more pronouncedly at the low-energy threshold for $q<2k_F$.
This is again in agreement with the nonlinear TLL theory, predicting a larger negative
exponent $\mu(q)$.

The dynamical structure factors in panels (a), (b) are also in qualitative agreement with
numeric calculations for the super Tonks-Girardeau gas, for which $0.4 \lesssim K_L < 1$ \cite{Panfil2013}; in fact we verified that the spectra shown in Ref.~\cite{Panfil2013} manifest
a low-energy support at positive energy which is compatible with Eq. \eqref{eq:lowenthr}, up to $\rho a\simeq 0.1$ (even though one should remark that a negative-frequency component is also present due to the excited nature of the super Tonks-Girardeau state).

The spectra in Figure \ref{fig:spectra} show that the Feynman approximation
\begin{equation}
\hbar \omega_{FA}(q) = \frac{\hbar^2 q^2}{2m S(q)}
\end{equation}
breaks down beyond $\frac{q}{2k_F} \simeq  0.1$. 
Interestingly, around $\frac{q}{2k_F} \simeq 0.1$ also the approximation of $\omega_{th}(q)$
with a linear function of $q$ ceases to be adequate.
The simultaneous appearance of nonlinear terms in $ \omega_{th}(q)$ and corrections to the 
Feynman approximation in $S(q,\omega)$ are in fact deeply related phenomena, as explained by 
the nonlinear TLL theory.

When $K_L<1/2$, Eq. \eqref{eq:skpeaks} indicates that a peak manifests in the static structure factor at $q=2k_F$. This change in behavior is
also reflected in $S(q,\omega)$, as shown in panels (c), (d). The spectral weight concentrates close to the
lower branch $\omega^*_-(q)$ of the particle-hole band, in a region of dense spectral weight
that we call lower mode following \cite{Bertaina2016}, where a similar behavior was observed in 1D ${}^4$He at high density.
In both HRs and ${}^4$He, above the lower mode stretches a high-energy structure gathering a
smaller fraction of spectral weight.
Such high-energy structure has a minimum at $q = 2k_F$ close to the free-particle energy 
$\energycommand = 4 E_F$, and is symmetric around $q=2k_F$, (see panel (d) in Figure \ref{fig:spectra}).

Panel (d) also shows that, as $K_L$ decreases below $1/2$, the support of $S(q,\omega)$
extends below $\omega^*_-(q)$ for $q>2k_F$, still remaining above $\omega_{th}(q)$.
In the high-density regime $\rho a \geq 0.642$, such region of the momentum-energy plane
hosts some of the most remarkable properties of $S(q,\omega)$ as it can be read from panel
(e), where the shape of $S(q,\omega)$ changes considerably for $2k_F < q< 4k_F$.

For $q < 2.8 k_F$, the spectral weight concentrates in a narrow region of the momentum-energy
plane that gradually departs from the low-energy threshold.
For $2.8 k_F < q < 3.6 k_F$, $S(q,\omega)$ suddenly and considerably broadens and flattens.
Finally, for $3.6 k_F < q < 4   k_F$, the spectral weight again concentrates close to 
the low-energy threshold.
This highly non-trivial behavior is in qualitative agreement with the nonlinear Luttinger 
liquid theory, predicting a negative exponent for $q<\mathcal{Q}_1$, where $\frac{\mathcal{Q}_1}{2 k_F} = \frac{1}{a\rho} = 1.558$.
At $q = \mathcal{Q}_1$ the non-linear Luttinger liquid theory predicts a flat dynamical
structure factor close to the low-energy threshold, in agreement with an exact prediction 
by F. Mazzanti {\em{et al.}} \cite{Mazzanti2008a} and with our observations for the 1D HR system, but also for a 1D system of $^4$He atoms\cite{Bertaina2016}.
Beyond $\mathcal{Q}_1$ the non-linear Luttinger liquid theory predicts a positive 
exponent and we observe the spectral weight concentrating close to the low-energy
threshold, as for $q< 2 k_F$. Notice that, although the condition for having quasi-Bragg peaks is $K_L<1/2$, the first special momentum with flat spectrum appears only for $\rho a > 1/2$, namely $K_L<1/4$, since one must have $\mathcal{Q}_1=2\pi/a < 4\pi \rho$.

It is well-known \cite{Mazzanti2008b,Mazzanti2008a} that, in the high-density
regime $K_L \ll 1$, HRs show up a packing order leading to a quasi-solid phase, 
crystallization being prohibited by the reduced dimensionality and by the range of the
interaction.
The emergence of the quasi-solid phase is signaled by the peaks of the static structure
factor, that approach the linear growth with the system size, peculiar prerogative of
the Bragg peaks, only in the $\rho a \to 1$ limit.

\begin{figure}
\centering
\includegraphics[width=0.48\textwidth]{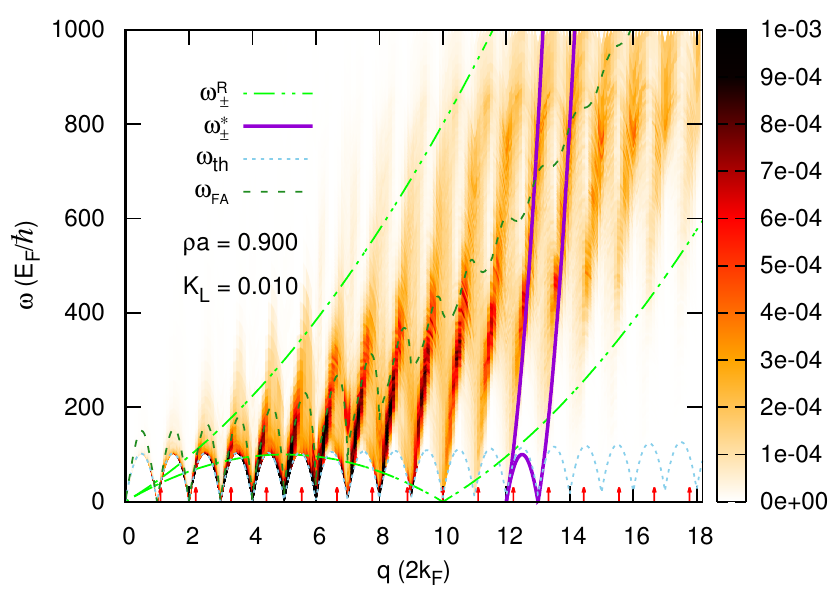}
\caption{(color online) Color map of the dynamic structure factor at $\rho a = 0.900$, on a much larger momentum and frequency scales than in Fig.~\ref{fig:spectra}(f). Units and reference curves are the same as in Fig.~\ref{fig:spectra}.
The special wavevectors $\mathcal{Q}_n$ (red arrows) are shown by long (red) arrows. The role of both the HR renormalized particle-hole frequencies (solid curves) and of the reduced-volume ideal Fermi gas band (dot-dashed curves) are shown.}
\label{fig:spectrum900}
\end{figure}

As far as $S(q,\omega)$ is concerned, the increase of the density transfers spectral weight to 
the low-energy threshold at wave vectors $q = 2k_F, 4k_F$.
We interpret this phenomenon as revealing that $S(q,\omega)$ is gradually approaching 
translational invariance $q \to q + 2 k_F$ in the variable $q$.
This conjecture is corroborated by the observation that, as the density is further
increased, the exponent $\mu_n(q)$ pointwise converges to the periodic function:
\begin{equation} \label{eq:munll}
\mu_n(q) = -2\left( \frac{q}{2 k_F} - n \right) \left( \frac{q}{2 k_F} - (n+1) \right) 
\end{equation}
with $2 n k_F \leq q \leq 2 (n+1) k_F$ (illustrated in Figure \ref{fig:mu_hr}). In this respect, it is interesting to observe
panel (f) of Figure \ref{fig:spectra}, showing the dynamical structure factor of HRs at
$\rho a = 0.900$. For $q< 8 k_F$, the spectral weight almost always concentrates around $\omega_{th}(q)$,
except in the small ranges of wavevectors $2n k_F<q<\mathcal{Q}_n$.
This makes $S(q,\omega)$ resemble the dispersion relation of longitudinal phonons of a 
monoatomic chain, in this range of momenta. However, the non-commensurability of $\mathcal{Q}_1$ with $2k_F$ renders the spectrum only quasi-periodic,
a behavior which is more and more manifest at higher momenta. To analyze this intriguing regime in more detail, we have reconstructed the spectra at $\rho a=0.900$ up to $q=38k_F$. The results are shown in Fig. \ref{fig:spectrum900} and indicate a crucial role of the reduced-size ideal Fermi gas in drawing large-scale momentum and frequency boundaries for the HR spectrum. If we define the density $\rho^\prime=\rho/(1-\rho a)$, we observe that above $q=20k_F=2\pi \rho^\prime$, namely twice the Fermi momentum of the reduced-size IFG, the spectrum is never peaked along the low-energy HR threshold, however it presents a stripe structure, repeating the Lieb-I mode plus multiple umklapp excitations, and becoming again flat at momenta $\mathcal{Q}_n$. At the level of accuracy of our GIFT reconstructions, the stripes are bounded by the particle-hole band of the reduced-size IFG
\begin{equation}
\hbar\omega_\pm^R(q)= \frac{\hbar^2}{2m} \left|2 \pi \rho^\prime  q \pm q^2\right|\quad. 
\end{equation}
To corroborate this observation, we find that the special momenta $\mathcal{Q}_n$ also analytically correspond to the crossings of the lower reduced-size IFG boundary and the HR threshold (for $q< 2\pi\rho^\prime$) or the HR repeated Lieb-I modes ($q> 2\pi\rho^\prime$). 
It is tempting to conclude that the HR spectrum can be almost completely described by the synergy of two rescaled ideal Fermi gases: one with the same density, but renormalized mass $m^*=m K_L$, the other with the same mass, but increased density $\rho^\prime=\rho/K_L^{1/2}$.

Only in the unphysical $\rho a \to 1$ limit, therefore, $S(q,\omega)$ would attain the translational 
invariance observed for instance in half-filled Hubbard chains with strong on-site repulsion 
\cite{Pereira2012,Roux2013}.

\begin{figure}
\centering
\includegraphics[width=0.42\textwidth]{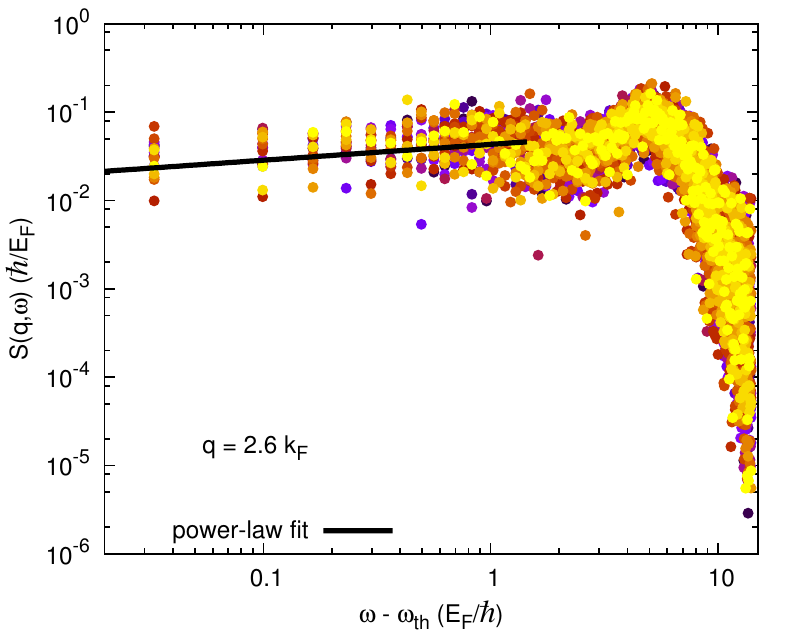}
\caption{(color online) Power-law fit of reconstructed spectra for $\rho a=0.700$ and $q=2.6 k_F$. The fitted exponent is $\mu=-0.18(3)$, which is compatible with the analytical prediction \eqref{eq:munll} that yields $\mu=-0.196$.}
\label{fig:exponent65}
\end{figure}

\begin{figure}
\centering
\includegraphics[width=0.42\textwidth]{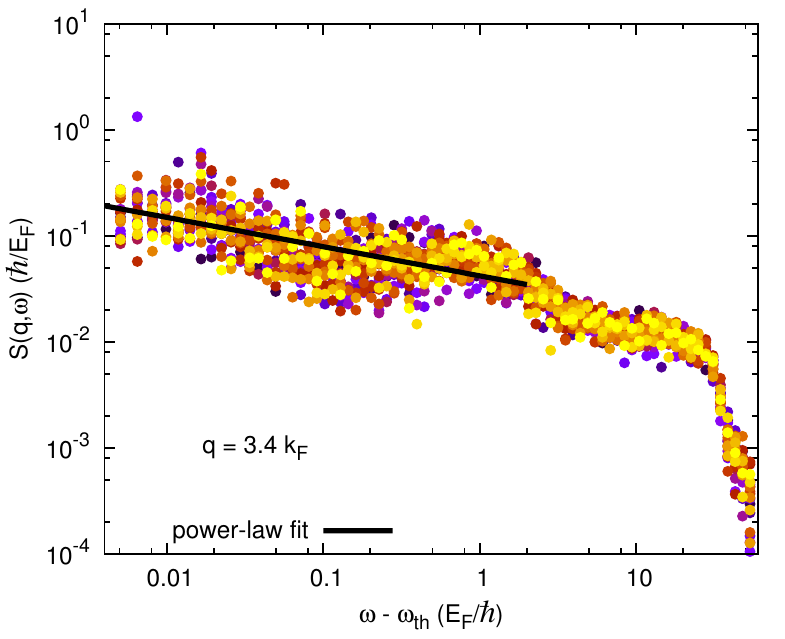}
\caption{(color online) Power-law fit of reconstructed spectra for $\rho a=0.700$ and $q=3.4 k_F$. The fitted exponent is $\mu=0.28(3)$, which is compatible with the analytical prediction \eqref{eq:munll} that yields $\mu=0.308$.}
\label{fig:exponent85}
\end{figure}

Finally, in discussing Figure \ref{fig:spectra}, we stressed in multiple occasions that low-energy 
properties of $S(q,\omega)$ are captured by the nonlinear TLL theory. Following the approach of \cite{Bertaina2016}, in Figures \ref{fig:exponent65} and \ref{fig:exponent85} we show that the agreement is quantitative, focusing on density $\rho a=0.700$ and two momenta close to $\mathcal{Q}_1$, representative of negative and positive power-law exponents, and fitting multiple spectral reconstructions with the functional form given by Eq. \eqref{eq:powerlaw}. The agreement with the analytical expression for the power-law exponent \eqref{eq:mu_hr} is good, even though the accuracy is strongly dependent on the quality of the original $F(q,\tau)$ and it is particularly delicate to fit the spectra when the spectrum goes to zero close to the threshold.

\section{Conclusions and Outlooks}
\label{sec:conc}

We have computed the zero-temperature dynamical structure factor of one-dimensional
hard rods by means of state-of-the-art QMC and analytic continuation techniques.
By increasing the rod length, the dynamical structure factor reveals a transition
from the Tonks-Girardeau gas, to a super Tonks-Girardeau regime and finally to a quasi-solid regime. 

The low-energy properties of the dynamical structure factor are in qualitative agreement
with the nonlinear LL theory.
However, the methodology provides a quantitative estimation of the dynamical structure
factor also in the high-energy regime, lying beyond the reach of LL theories.
Our study reveals strong similarities between the dynamical structure factor of
HRs and 1D ${}^4$He at linear densities $\rho \ge 0.150$~\AA$^{-1}$ \cite{Bertaina2016}, extending to the high-energy regime (well above the low-energy threshold).
In particular, both systems show a flat dynamical structure factor in correspondence of
the wavevectors $\mathcal{Q}_n$, in agreement with a previous theoretical prediction 
\cite{Mazzanti2008a}, and feature a high-energy structure overhanging the lower mode around 
the umklapp point $q=2k_F$. We have also unveiled a peculiar structure of the spectrum in the high-density and high-momentum regime,
which can be described in terms of the particle-hole boundaries of two renormalized ideal Fermi gases. 

At this point we want to remark that an intriguing feature of 1D $^4$He (and arguably of all 1D quantum liquids which admit a two-body bound state) is that its Luttinger parameter $K_L$ spans all 
positive values $0< K_L <\infty$ as the linear density is increased \cite{Bertaina2016}, not only the $K_L\le 1$ regime, as in the case of HR or dipolar systems \cite{Citro2007}. This is due to the attractive tail of the interaction potential, which dominates at low density. Such feature has also to be contrasted to the repulsive Lieb-Liniger model, for which one obtains $1\le K_L \le \infty$ by tuning the interaction strength. Finally also the Calogero-Sutherland model reproduces all possible $K_L$, however by tuning interaction only \cite{Astrakharchik2006}.

We hope the present work will encourage further experimental research in 1D systems with 
volume exclusion effects, and theoretical investigation of the HR model. Our results may be relevant also for the linear response dynamics of resonant Rydberg gases in 1D configurations \cite{Petrosyan2013}. On the theoretical side, possible directions for further developments might be the 
alternative calculations of $S(q,\omega)$, based e.g. on the VMC evaluation of the matrix 
elements of $\rho_q$, and/or the calculation of finite-temperature equilibrium and dynamical 
properties.

\begin{acknowledgments}
We acknowledge useful discussions with G. Astrakharchik. We thank M. Panfil and co-authors for providing us with their data on the super Tonks-Girardeau gas \cite{Panfil2013}.
We acknowledge the CINECA awards $\text{IsC29-SOFTDYN}$, $\text{IsC32-EleMEnt}$ and CINECA and Regione Lombardia LISA award $\text{LI05p-PUMAS}$, for the availability of high-performance computing resources and support.
We also acknowledge computing support from the computational facilities at the College of William and Mary
and at the Physics Department of the University of Milan.
M.M. and E.V. acknowledge support from the Simons Foundation and NSF (Grant no. DMR-1409510). M.R. acknowledges the EU QUIC project for fundings.
\end{acknowledgments}

\bibliographystyle{unsrt}

\begin{thebibliography}{99}
\bibitem{Giamarchi2004} T. Giamarchi, {\emph{Quantum Physics in One Dimension}}, Oxford University Press (2004)
\bibitem{Cazalilla2011} M.A. Cazalilla, R. Citro, T. Giamarchi, E. Orignac and M. Rigol, {\em{Rev. Mod. Phys.}} {\bf{83}}, 1405 (2011)
\bibitem{Imambekov2012} A. Imambekov, T. L. Schmidt and L. I. Glazman, {\em{Rev. Mod. Phys.}} {\bf{84}}, 1253 (2012)
\bibitem{HMW} P.C. Hohenberg, {\em{Phys. Rev.}} {\bf{158}}, 383 (1967)
\bibitem{Mermin1966} N. D. Mermin and H. Wagner, {\em{Phys. Rev. Lett.}} {\bf{17}}, 1133 (1966)
\bibitem{Coleman1973} S. Coleman, {\em{Commun. Math. Phys.}} {\bf{31}}, 259 (1973)
\bibitem{Girardeau1960} M. Girardeau, {\emph{J. Math. Phys.}} {\bf{1}}, 516 (1960)
\bibitem{Pitaevskii1993} L. Pitaevskii and S. Stringari, {\em{Phys. Rev. B}} {\bf{47}}, 10915 (1993)
\bibitem{Vignale2005} G. F. Giuliani and G. Vignale, {\em{Quantum Theory of the
Electron Liquid}}, Cambridge University Press (2005)
\bibitem{Nagamiya1940} T. Nagamiya,
                       {\em Proc. Phys. Math. Soc. Jpn.} {\bf 22}, 705 (1940)
\bibitem{LiebLiniger1963} E. H. Lieb and W. Liniger, {\em{Phys. Rev.}} {\bf{130}}, 1605 (1963)
\bibitem{Lieb1963} E.H. Lieb, {\em{Phys. Rev.}} {\bf{130}}, 1616 (1963).
\bibitem{Sutherland1971b} B. Sutherland, {\em{J. Math. Phys.}} {\bf{12}}, 251 (1971)
\bibitem{Sutherland1971} B. Sutherland {\em Phys. Rev. A} {\bf 4}, 2019 (1971)
\bibitem{Sutherland1988} B. Sutherland, {\em{Phys. Rev. B}} {\bf{38}}, 6689 (1988)
\bibitem{expgen1} P . Paredes, {\em{Nature}} {\bf{429}}, 277 (2004)
\bibitem{expgen2} T. Kinoshita, T. Wenger and D. S. Weiss, {\em{Science}} {\bf{305}}, 1125 (2004)
\bibitem{expgen3} E. Haller, M. Gustavss, M. J. Mark, J. G. Danzl, R. Hart, G. Pupillo and H. C. Negerl, {\em{Science}} {\bf{325}}, 1224 (2009)
\bibitem{expdyn1} H. P. Stimming, N. J. Mauser, J. Schmiedmayer and I. E. Mazets, {\em{Phys. Rev. Lett.}} {\bf{105}}, 015301 (2010)
\bibitem{expdyn2} M. Kuhnert, R. Geiger, T. Langen, M. Gring, B. Rauer, T. Kitagawa, E. Demler, D. Adu Smith and J. Schmiedmayer, {\em{Phys. Rev. Lett.}} {\bf{110}}, 090405 (2013)
\bibitem{expdyn3} S. Hofferberth, I. Lesanovsky, B. Fischer, T. Schumm and J. Schmiedmayer, {\em{Nature}} {\bf{449}}, 324 (2007)
\bibitem{expdyn4} T. Langen, R. Geiger, M. Kuhnert, B. Rauer and J. Schmiedmayer, {\em{Nature Physics}} {\bf{9}}, 640 (2013)
\bibitem{expsol1} E. Witkowska, P. Deuar, M. Gajda and K. Rzazewski, {\em{Phys. Rev. Lett.}} {\bf{106}}, 135301 (2011)
\bibitem{expsol2} T. Karpiuk, P. Deuar, P. Bienias, E. Witkowska, K. Pawlowski, M. Gajda, K. Rzazewski and M. Brewczyk, {\em{Phys. Rev. Lett.}} {\bf{109}}, 205302 (2012)
\bibitem{expfm1} V. Guarrera, D. Muth, R. Labouvie, A. Vogler, G. Barontini, M. Fleischhauer and H. Ott, {\em{Phys. Rev. A}} {\bf{86}}, 021601 (2012)
\bibitem{expfm2} J.P. Ronzheimer, M. Schreiber,S. Braun,S.S. Hodgman,S. Langer,I.P. McCulloch,F. Heidrich-Meisner,I. Bloch,U. Schneider, {\em{Phys. Rev. Lett.}} {\bf{110}}, 205301 (2013)
\bibitem{Fabbri2015} N. Fabbri, M. Panfil, D. Cl\'ement, L. Fallani, M. Inguscio, C. Fort, and J. S. Caux {\em{Phys. Rev. A}} {\bf{91}}, 043617 (2015)
\bibitem{Bertaina2016} G. Bertaina, M. Motta, M. Rossi, E. Vitali and D. E. Galli, {\em{Phys. Rev. Lett.}} {\bf{116}}, 135302 (2016)
\bibitem{Pearce2005} J. V. Pearce, M. A. Adams, O. E. Vilches, M. R. Johnson and H. R. Glyde, {\em{Phys. Rev. Lett.}} {\bf{95}}, 185302 (2005)
\bibitem{Mercedes2001} M. Mercedes Calbi, M. W. Cole, S. M. Gatica, M. J. Bojan and G. Stan, {\em{Rev. Mod. Phys.}} {\bf{73}}, 857 (2001)
\bibitem{Mazzanti2008b} F. Mazzanti, G. E. Astrakharchik, J. Boronat and J. Casulleras, 
                       {\em Phys. Rev. A} {\bf 77}, 043632 (2008)
\bibitem{Mazzanti2008a} F. Mazzanti, G. E. Astrakharchik, J. Boronat and J. Casulleras, 
                       {\em Phys. Rev. Lett.} {\bf 100}, 020401 (2008)
\bibitem{Waals1910} J. D. Van der Waals, {\em{The equation of state for gases and liquids}}, Nobel Lectures in Physics 254 (1910)
\bibitem{Jeans1916} J. H. Jeans, {\em{The Dynamical Theory of Gases}}, Cambridge University Press (1916)
\bibitem{Tonks1936} L. Tonks, {\em{Phys. Rev.}} {\bf{50}}, 955 (1936)
\bibitem{Hove1952} B. R. A. Nijboer and L. Van Hove, {\em{Phys. Rev.}} {\bf{85}}, 777 (1952)
\bibitem{Krotscheck1999} E. Krotscheck, M.D. Miller, and J. Wojdylo, {\em{Phys. Rev. B}} {\bf{60}}, 13028 (1999).
\bibitem{Sarsa2000}A. Sarsa, K. E. Schmidt, and W. R. Magro, {\em{J. Chem. Phys.}}
{\bf{113}}, 1366 (2000)
\bibitem{Galli2003} D. E. Galli and L. Reatto, {\em{Mol. Phys.}} {\bf{101}}, 1697 (2003)
\bibitem{Patate} M. Rossi, M. Nava, L. Reatto, and D.E. Galli, {\em{J. Chem. Phys.}} {\bf 131}, 154108 (2009)
\bibitem{Vitali2010} E. Vitali, M. Rossi, L. Reatto and D. E. Galli, {\em{Phys.
Rev. B}} {\bf{82}}, 174510 (2010)
\bibitem{Cowley1971} R.A. Cowley and A.D.B. Woods, {\em{Can. J. Phys.}} {\bf{49}}, 177 (1971).
\bibitem{Beauvois2016} K. Beauvois, C.E. Campbell, J. Dawidowski, B. F\aa{}k, H. Godfrin, E. Krotscheck, H.-J. Lauter, T. Lichtenegger, J. Ollivier, and A. Sultan, {\em{Phys. Rev. B}} {\bf{94}}, 024504 (2016).
\bibitem{Ozeri2005} R. Ozeri, N. Katz, J. Steinhauer, and N. Davidson, {\em{Rev. Mod. Phys.}} {\bf{77}}, 187 (2005).
\bibitem{Ha2015} L.-C. Ha, L.W. Clark, C.V. Parker, B.M. Anderson, and C. Chin, {\em{Phys. Rev. Lett.}} {\bf{114}}, 055301 (2015).
\bibitem{Landig2015} R. Landig, F. Brennecke, R. Mottl, T. Donner, and T. Esslinger, {\em{Nat. Commun.}} {\bf{6}}, (2015).
\bibitem{Tomonaga1950} S. Tomonaga, {\em{Prog. Theor. Phys.}} {\bf{5}}, 544 (1950)
\bibitem{Luttinger1963} J. M. Luttinger, {\em{J. Math. Phys.}} {\bf{4}}, 1154 (1963)
\bibitem{Mattis1965} D. C. Mattis and E. H. Lieb, {\em{J. Math. Phys.}} {\bf{6}}, 304 (1965)
\bibitem{Haldane1983} F. D. M. Haldane, {\em{Phys. Rev. Lett.}} {\bf{47}}, 1840 (1981)
\bibitem{Haldane1983b} F. D. M. Haldane, {\em{Phys. Rev. Lett.}} {\bf{48}}, 569 (1982)
\bibitem{Imambekov2009} A. Imambekov and L. I. Glazman, {\em{Phys. Rev. Lett.}} {\bf{102}}, 126405 (2009)
\bibitem{Bethe1931} H. Bethe, {\em Zeit. Phys.} {\bf 71}, 205 (1931)
\bibitem{Panfil2013} M. Panfil, J. De Nardis and J.-S. Caux, {\em{Phys. Rev. Lett.}} {\bf{110}}, 125302 (2013)
\bibitem{Astrakharchik2005} G. E. Astrakharchik, J. Boronat, J. Casulleras and S. Giorgini, {\em{Phys. Rev. Lett.}} {\bf{95}}, 190407 (2005)
\bibitem{Batchelor2005} M. T. Batchelor, M. Bortz, X. W. Guan and N. Oelkers, {\em{J. Stat. Mech.}} {\bf{L10001}} (2005)
\bibitem{Tempfli2008} E. Tempfli, S. Z\"ollner and P. Schmelcher, {\em{New J. Phys.}} {\bf{10}}, 103021 (2008)
\bibitem{Haller2009} E. Haller, M. Gustavsson, M.J. Mark, J.G. Danzl, R. Hart, G. Pupillo, and H.-C. Nägerl, {\em Science} {\bf{325}}, 1224 (2009).
\bibitem{CastroNeto1994} A.H. Castro Neto, H.Q. Lin, Y.-H. Chen, and J.M.P. Carmelo, {\em{Phys. Rev. B}} {\em{50}}, 14032 (1994).
\bibitem{Cherny2011} A. Y. Cherny, J. S. Caux and J. Brand {\em{Frontiers of Physics}} {\bf{7}}, 54 (2012)
\bibitem{Brand2016} S.S. Shamailov and J. Brand, {\em{New J. Phys.}} {\bf{18}}, 75004 (2016)
\bibitem{Caux2006} J. S. Caux and P. Calabrese, {\em{Phys. Rev. A}} {\bf{74}}, 031605 (2006)
\bibitem{Meinert2015} F. Meinert, M. Panfil, M. J. Mark, K. Lauber, J.-S. Caux and H.-C. N\"agerl, {\em{Phys. Rev. Lett.}} {\bf{115}}, 085301 (2015)
\bibitem{DePalo2008} S. De Palo, E. Orignac, R. Citro, and M.L. Chiofalo, {\em{Phys. Rev. B}} {\bf{77}}, 212101 (2008)
\bibitem{Nava2013} M. Nava, D.E. Galli, S. Moroni and E. Vitali, {\em{Phys. Rev. B}} {\bf{87}}, 144506 (2013)
\bibitem{Arrigoni2013} F. Arrigoni, E. Vitali, D. E. Galli and L. Reatto, {\em{Low Temp. Phys.}} {\bf{39}}, 793 (2013)
\bibitem{Rota2013} R. Rota, F. Tramonto, D.E. Galli and S. Giorgini, {\em{Phys. Rev. B}} {\bf{88}}, 214505 (2013)
\bibitem{Holzmann2003} M. Holzmann, D. M. Ceperley, C. Pierleoni and K. Esler, {\em{Phys. Rev. E}}
{\bf{68}}, 046707 (2003)
\bibitem{Kalos1986}  M. H. Kalos and P. A. Whitlock {\em{Quantum Monte Carlo}}, in {\em{Monte Carlo Methods}}, Wiley (1986)
\bibitem{Toulouse2007} J. Toulouse and C. J. Umrigar, {\em{J. Chem. Phys.}} {\bf{126}}, 084102 (2007)
\bibitem{Motta2015} M. Motta, G. Bertaina, D. E. Galli and E. Vitali, {\em{Comp. Phys. Comm.}}
                    {\bf{190}}, 62 (2015)
\bibitem{Metropolis1953} N. Metropolis, A. W. Rosenbluth, M. N. Rosenbluth, A. H. Teller and E. Teller, {\em{J. Chem. Phys.}} {\bf{21}} 1087 (1953)
\bibitem{PP} D. Ceperley, {\em{Rev. Mod. Phys.}} {\bf{67}}, 279 (1995)
\bibitem{CaoBerne1992} J. Cao and B.J. Berne, {\em{J. Chem. Phys.}} {\bf{97}}, 2382 (1992)
\bibitem{Gordillo2012} F. De Soto and M.C. Gordillo, {\em{Phys. Rev. A}} {\bf{85}}, 013607 (2012)
\bibitem{Tarantola2006} A. Tarantola, {\em{Nature Physics}} {\bf{2}}, 492 (2006)
\bibitem{MEM1} R. N. Silver, J.E. Gubernatis, D. S. Sivia and M. Jarrell, {\em{Phys. Rev. Lett.}} {\bf{65}}, 496 (1990)
\bibitem{MEM2} M. Jarrell and J.E. Gubernatis, {\em{Phys. Rep.}} {\bf{269}}, 133 (1996)
\bibitem{MEM3} M. Boninsegni, and D.M. Ceperley, {\em{J. Low Temp. Phys.}} {\bf{104}}, 339 (1996)
\bibitem{MEM4} A. Roggero, F. Pederiva and G. Orlandini {\em{Phys. Rev. B}} {\bf{88}}, 094302 (2013)
\bibitem{Mishchenko2000} A.S. Mishchenko, N.V. {Prokof’ev}, A. Sakamoto, and B.V. Svistunov, {\em{Phys. Rev. B}} {\bf{62}}, 6317 (2000).
\bibitem{Overpress} M. Rossi, E. Vitali, L. Reatto and D.E. Galli, {\em{Phys. Rev. B}} {\bf{85}}, 014525 (2012)
\bibitem{Anisotropic} M. Nava, D.E. Galli, M.W. Cole, L. Reatto, {\em{J. Low Temp. Phys.}} {\bf 171}, 699 (2013)
\bibitem{Molinelli2016} S. Molinelli, D. E. Galli , L. Reatto and M. Motta, {\em{J. Low Temp. Phys.}} DOI: 10.1007/s10909-016-1628-3 (2016)
\bibitem{Saccani2011} S. Saccani, S. Moroni, E. Vitali, and M. Boninsegni, {\em{Mol. Phys.}} {\bf{109}}, 2807 (2011).
\bibitem{Saccani2012} S. Saccani, S. Moroni, and M. Boninsegni, {\em{Phys. Rev. Lett.}} {\bf{108}}, 175301 (2012).
\bibitem{Vitali2016} E. Vitali, H. Shi, M. Qin, and S. Zhang, {\em{Phys. Rev. B}} {\bf{94}}, 085140 (2016).
\bibitem{Teruzzi2016} M. Teruzzi, D.E. Galli, and G. Bertaina, arXiv:1607.05308 (2016).
\bibitem{Astrakharchik2014} G. E. Astrakharchik and J. Boronat, {\em{Phys. Rev. B}} {\bf{90}}, 235439 (2014)
\bibitem{Roux2013} G. Roux, A. Minguzzi and T. Roscilde, {\em{New J. Phys.}} {\bf{15}}, 055003 (2013)
\bibitem{Pereira2012} R. G. Pereira, K. Penc, S. R. White, P. D. Sacramento and J. M. P. Carmelo, {\em{Phys. Rev. B}} {\bf{85}}, 165132 (2012)
\bibitem{Citro2007} R. Citro, E. Orignac, S. De Palo, and M.L. Chiofalo, {\em{Phys. Rev. A}} {\bf{75}}, 051602 (2007)
\bibitem{Astrakharchik2006} G.E. Astrakharchik, D.M. Gangardt, Y.E. Lozovik, and I.A. Sorokin, {\em{Phys. Rev. E}} {\bf{74}}, 021105 (2006)
\bibitem{Petrosyan2013} D. Petrosyan, M. H\"oning, and M. Fleischhauer, {\em{Phys. Rev. A}} {\bf{87}}, 053414 (2013)

\end{thebibliography}

\end{document}